\documentclass[cernpreprint,english]{na61doc}

\usepackage{amsmath}
\usepackage{mathtools}
\usepackage{amssymb}
\usepackage{commath}
\usepackage{rviewport}
\usepackage[nottoc]{tocbibind}
\usepackage{lineno}
\usepackage{placeins}
\usepackage{array,makecell,bigdelim,booktabs}
\usepackage[separate-uncertainty=true, table-align-uncertainty=true]{siunitx}

\usepackage{colortbl}
\definecolor{darkred}{rgb}{0.5,0,0}
\definecolor{darkblue}{rgb}{0,0,0.5}
\definecolor{firebrick}{rgb}{0.75,0.125,0.125}
\definecolor{darkgreen}{rgb}{0,0.5,0}
\usepackage{multirow}
\usepackage{comment}
\usepackage{diagbox}

\usepackage[colorlinks=true,linkcolor=firebrick,citecolor=darkgreen,urlcolor=darkblue]{hyperref}
\usepackage[capitalise]{cleveref}
\usepackage{booktabs}
\usepackage{diagbox}
\usepackage{cite}
\usepackage{upgreek}
\DeclareMathOperator{\erf}{erf}

\newcommand{\iso}[2]{$^{#1}$#2}

\newcommand{\GeVc}{\ensuremath{\text{GeV}/c}\xspace}
\newcommand{\GVc}{\ensuremath{\text{GV}/c}\xspace}

\newcommand{\AGeVc}{\ensuremath{A\,\text{GeV}/c}\xspace}

\newcommand{\cmsq}{\ensuremath{\text{cm}^2}\xspace}

\newcommand{\cm}{\ensuremath{\text{cm}}\xspace}

\newcommand{\dd}{\ensuremath{\mathrm{d}}\xspace}

\newcommand{\dedx}{\ensuremath{\dd E/\dd x}\xspace}
\newcommand{\tof}{\ensuremath{\textup{\emph{tof}}}\xspace}

\def\sci#1#2{#1{\times}10^{#2}}

\newcommand{\Ctwelve}{\ensuremath{\text{\iso{12}{C}}}\xspace}

\newcommand{\ZsqGTPC}{\ensuremath{Z^2_\text{GTPC}}\xspace}
\newcommand{\ZsqVTPCtwo}{\ensuremath{Z^2_\text{VTPC-2}}\xspace}

\def\delx{\ensuremath{\Delta x}\xspace}
\def\dely{\ensuremath{\Delta y}\xspace}
\def\Zsq{\ensuremath{Z^2}\xspace}
\def\b{`b'\xspace}
\def\f{`f'\xspace}
\newcommand{\Ps}[2]{\ensuremath{P_{\text{#1}\to\text{#2}}}\xspace}
\newcommand{\PP}[3]{\ensuremath{P^\text{#1}_{\text{#2}\to\text{#3}}}\xspace}

\ShineTitle{Measurement of the mass-changing, charge-changing and production cross sections of \iso{11}{C}, \iso{11}{B} and \iso{10}{B} nuclei in \Ctwelve+p interactions at 13.5\,\GeVc per nucleon.}

\PreprintIdNumber{CERN-EP-2024-278}

\ShineJournal{PRC}
\ShineAbstract{ %
We report results from a 2018 pilot run to study the feasibility of nuclear fragmentation measurements with the NA61/SHINE experiment at the CERN SPS.
These results are important for the interpretation of the production of light secondary cosmic-ray nuclei (Li, Be, and B) in the Galaxy.
The specific focus here is on cross sections important for the production of boron in the Galaxy from the interactions of \Ctwelve nuclei with hydrogen in the interstellar medium, including the contribution from the decay of the short-lived \iso{11}{C} fragments.
The data were taken with the secondary \Ctwelve beam at beam momentum of 13.5\,\GeVc per nucleon and two fixed targets, polyethylene (CH$_2$) and graphite (C), from which we derive the cross sections of carbon on hydrogen.
We present the measurement of the fragmentation cross sections of \iso{11}{C}, \iso{11}{B}, and \iso{10}{B} as well as the
mass- and charge-changing cross sections.
\\[10mm]
}
\begin{document}

\maketitle

\section{Introduction}
\label{intro}

Galactic Cosmic Rays (GCR) are mainly comprised of highly relativistic nuclei, and are categorized based on their origin as primary and secondary. While primary cosmic rays are accelerated at cosmic ray sources such as supernovae remnants, secondary cosmic rays are spallation products of primaries interacting with the interstellar medium (ISM). Studying propagation characteristics of GCRs is crucial in determining background contribution to dark-matter annihilation signals, estimating the size of the halo of the Galaxy, and provide insights into new physics by predicting flux of cosmic ray lithium and fluorine, which are expected to be of purely secondary origin~\cite{Maurin_2022,PhysRevLett.126.081102,Zhao_2023}. Two key inputs for modeling the propagation of GCR are the secondary-to-primary flux ratios arriving at Earth and the nuclear fragmentation cross section values.
The most studied ratio is that of the boron-to-carbon flux ratio, which is used to infer the total amount of matter traversed by the cosmic rays in the Galaxy~\cite{maurin02,strong07,tomassetti17,evoli18}.
Recent measurements by space-based cosmic ray detectors, like PAMELA, AMS-02, CALET, and  DAMPE~\cite{pamelaCollab,amsCollab,caletCollab,dampeCollab} show improved flux uncertainties, reaching levels ${<}5\%$.
On the other hand, current nuclear fragmentation cross section values are known with a precision not less than ${\sim}20\%$.
These uncertainties dominate the uncertainties in the CR propagation models~\cite{genolini2018,evoli2020,genolini2023}.

Cosmic boron is produced in spallation reactions of the primary cosmic ray nuclei like C, N, and O interacting with the interstellar hydrogen.
The ratio is then used to estimate the total amount of matter traversed by the primary cosmic rays in the Galaxy, and deduce transport parameters like the normalization of the diffusion coefficient.
Therefore, measuring fragmentation cross sections of specific reactions with a level of precision equivalent to the measured fluxes is crucial.
Moreover, the spallation of primary GCR nuclei also produces short-lived radionuclides, called ghosts, like \iso{11}{C}.
The \iso{11}{C} nucleus decays to the stable \iso{11}{B} via $\upbeta^+$ decay with a half-life of ${\sim}20$\,min, and adds to the total boron (\iso{10}{B}+\iso{11}{B}) production in the Galaxy.
Such reactions emphasize the significance of measuring isotope production cross sections as well, with precision much better than the current values.

Nuclear fragmentation on proton targets have been a topic of interest for many groups. The most studied reaction is C+p$\to$A+X utilizing various accelerator facilities available at different beam energies.
Advancements in accelerator and detector facilities and detection techniques have led to more sophisticated measurement methods, an example of which is the fixed target experimental facility of \NASixtyOne at CERN.
In this paper, we will discuss the results of nuclear fragmentation cross sections in the reaction \iso{12}{C}+p$\to$B+X, at 13.5~\AGeVc beam momentum, using two targets, polyethylene (CH$_2$) and graphite. based on a pilot run conducted with \NASixtyOne at the Super Proton Synchrotron (SPS).

The paper is organized as follows: a brief introduction of the theoretical framework of nuclear fragmentation and the relevant cross section quantities measured in this analysis is given in~\cref{sec:nucFrag}. The details of the \NASixtyOne experiment and pilot run on fragmentation are provided in~\cref{sec:exp}, followed by~\cref{sec:data} with the data selection procedure adopted in this work.
We discuss the actual measurement of the distribution of the nuclear fragments produced in the beam-target interactions in \cref{sec:frag}.
The mathematical formalism developed for computing the mass-, charge-, and boron production cross sections with the corrections applied to the measurements, are derived in \cref{sec:analysis}.
Our pilot results are compared to existing data sets and cross section parametrizations in \cref{sec:res}, and we provide a brief outlook on the future runs for fragmentation studies at \NASixtyOne in \cref{sec:con}.

\section{Nuclear Fragmentation}
\label{sec:nucFrag}

The total reaction cross section $\upsigma_\mathrm{tot}$ characterizes the interaction of the projectile and the target nuclei irrespective of the number of particles in the final state. It is defined as the sum of the inelastic and elastic cross sections as,
\begin{equation*}
	\upsigma_\mathrm{tot} = \upsigma_\mathrm{inel} + \upsigma_\mathrm{el}.
\end{equation*}
where $\upsigma_\mathrm{el}$ is an elastic scattering interaction where the identity and  total kinetic energy of the two nuclei are the same, before and after the interaction. Whereas, inelastic cross section denoted as $\upsigma_\mathrm{inel}$ describes an interaction leading to production of at least one new particle. It can be further expressed as a sum of two quantities as,
\begin{equation*}
	\upsigma_\mathrm{inel} = \upsigma_\mathrm{prod} + \upsigma_\mathrm{qela},
\end{equation*}
where $\upsigma_\mathrm{prod}$ is the production cross section and it describes interactions involving production of at least one new hadron, while $\upsigma_\mathrm{qela}$ is the quasi-elastic cross section and  quantifies processes wherein a nucleus disintegrates to form lighter fragments upon interacting with the target nucleus, without producing any new hadrons. A theoretical framework based on Glauber theory of scattering, describing relativistic heavy ion collisions is called the abrasion-ablation model~\cite{Glauber1959,Townsend1984}. In the abrasion stage, nucleons in the overlap region of the colliding nuclei are removed. These nucleons are called participants, which can further decay into secondary hadrons such as pions. In the following stage of ablation, the non-participant nucleons called spectators, form a pre-fragment in the excited state, which can then disintegrate into lighter fragments.

In this work, we focus on the measurement of the partial inelastic
scattering cross section leading to the production of a boron nucleus,
either in the direct reactions \iso{12}{C}+p$\to$\iso{11}{B}+X and
\iso{12}{C}+p$\to$\iso{10}{B}+X, or via the indirect channel
\iso{12}{C}+p$\to$\iso{11}{C}$\to$\iso{11}{B}+X. Here the last
reaction denotes the $\beta^+$-decay of \iso{11}{C} with a half-life
of $\sim$ 20 minutes, i.e.\ astrophysically short-lived, but stable in
the context of the measurement discussed here.  The sum of these three
reactions is referred to as {\itshape the boron production cross
  section} throughout this paper.

Moreover, we present measurements of the {\itshape mass-changing}
cross section, $\upsigma_\mathrm{\Delta A}$, and {\itshape
  charge-changing} cross section, $\upsigma_\mathrm{\Delta Z}$, in which
the projectile nucleus loses at least one nucleon or at least one
proton, respectively. The corresponding reactions are,
\begin{equation*}
   ^{A}_Z\text{P} + \text{T} \rightarrow \; ^{A^{\prime}}_{Z^{\prime}}\text{P}^{\prime} + X,
\end{equation*}
where P denotes the projectile nucleus, T is the target nucleus, P$^\prime$ is
the leading outgoing fragment in the projectile hemisphere after the interaction, and $\Delta A = A
- A^{\prime} > 0$ for a mass-changing reaction and $\Delta Z = Z -
Z^{\prime} > 0$ for a charge-changing reaction for P. The leading outgoing
fragment P$^\prime$ is the nucleus with the largest mass/charge among the
reaction products in the projectile hemisphere.

\section{Experimental Setup}
\label{sec:exp}

The nuclear fragmentation cross section measurements reported in this work were performed at the \NASixtyOne facility at CERN during a pilot run in 2018.
This section encompasses the relevant details of the \NASixtyOne detector systems and the pilot run.

\subsection{The NA61/SHINE facility}
\label{sec:na61}

\begin{figure}
    \includegraphics[width=\textwidth]{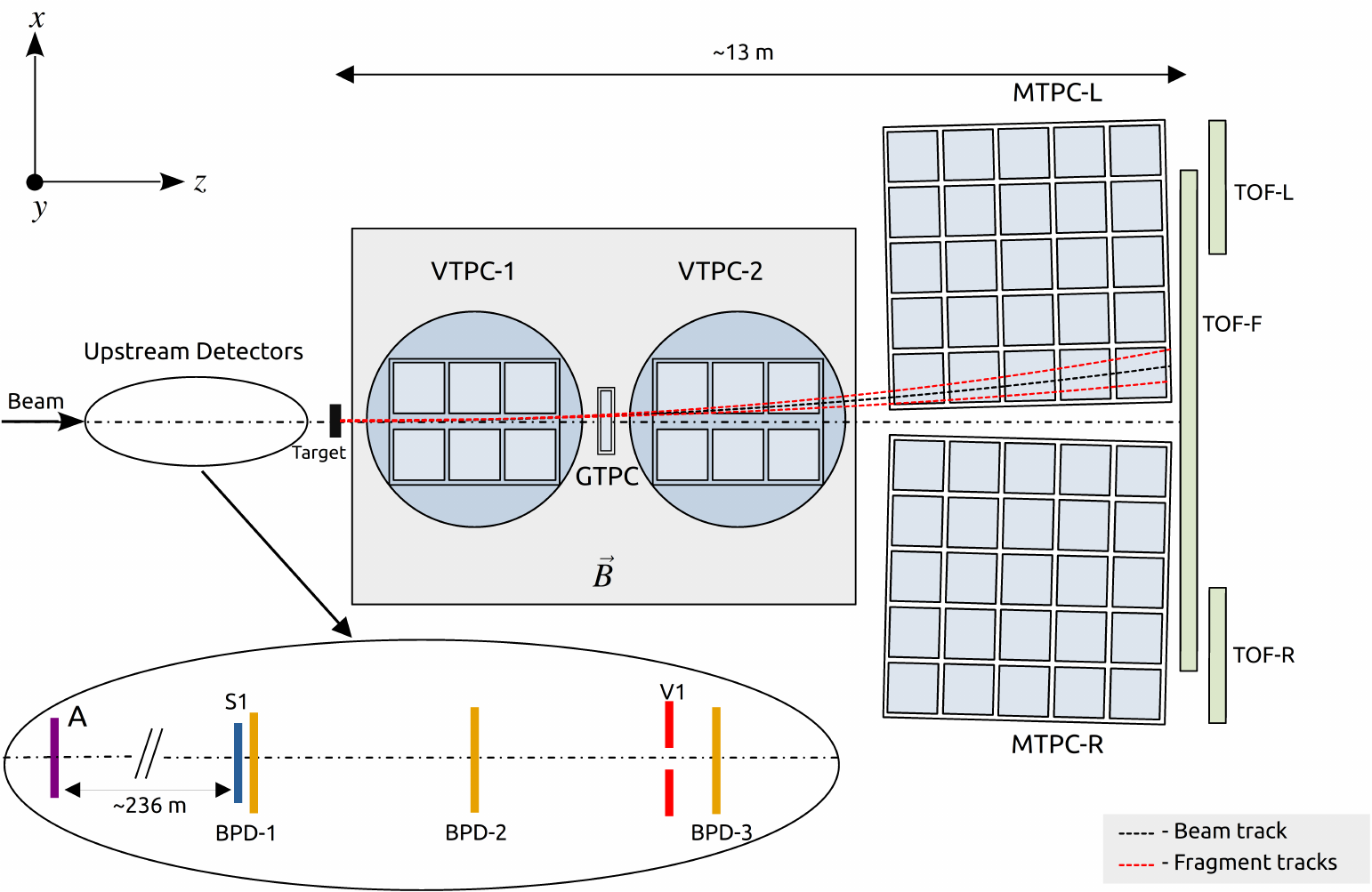}
    \caption{NA61/SHINE experiment depicting the layout of the detectors used during the pilot run for fragmentation in 2018. The black dashed line shows the trajectory of the beam particle deflected in the magnetic field of the Vertex magnet, while the red dashed lines represent the fragments produced from the beam-target interaction. The beamline detectors upstream of the target starting from the A-detector up to BPD-3, are shown in the inset. The TOF-F, TOF-R, and TOF-L downstream of the MTPC, are the time-of-flight detectors, and were not used for this measurement.}
    \label{fig:1}
\end{figure}

NA61/SHINE (SPS Heavy Ion and Neutrino Experiment) is located in the CERN North Area, on the H2 beam line of the Super Proton Synchrotron (SPS) and is the successor to the NA49 experiment~\cite{Abgrall:2014fa}.
It operates within a beam momentum range of approximately 13~\AGeVc to 158~\AGeVc for ions.
The physics programs of the experiment primarily focus on studying the nucleus-nucleus interactions in the interest of phase transitions in strong interactions and investigating the properties of neutrino beams produced in proton-nucleus collisions.
Along with these, its third main objective is related to cosmic ray measurements, such as the study of nuclear fragmentation reactions in the interest of GCR propagation and air-shower studies~\cite{Adhikary:2826863}.
The ion or proton beam set at a specific momentum is delivered by the SPS accelerator and transported to North Experimental Hall 1 (EHN1) via the H2 beamline, where NA61/SHINE is located.
Multiple sets of dipole and quadrupole magnets on the beamline are used to steer and focus the beam particles.

The main detectors of the facility used for the measurement primarily consist of three sets of Time Projection Chambers (TPCs) namely, the Vertex TPC (VTPC-1/2) placed inside two sets of superconducting magnets with a total bending power of 9\,Tm, the Gap TPC (GTPC) located in between the two VTPCs, and the large acceptance Main TPC placed on the left and right of the beam pipe (MTPC-L/R), as shown in \cref{fig:1}.
Each of the MTPCs has the dimensions $L{\times}W{\times}H = 390{\times}390{\times}180$\,cm$^3$, whereas the dimensions of each of the VTPCs is $L{\times}W{\times}H = 250{\times}200{\times}98$\,cm$^3$, and that of the GTPC is  $L{\times}W{\times}H=30\times81.5\times70$\,cm$^3$.
The gas mixture in the TPC systems is Ar/CO$_2$ in the ratio 90/10 for the VTPCs and the GTPC, and 95/5 for the MTPC.
The two superconducting magnets hosting the VTPCs deflect the charged particles produced in the beam-target interaction, depending on their rigidity\footnote{Rigidity of a particle with electric charge $Z$ moving perpendicular to magnetic field $B$ is a measure of its deflection and is defined as $R=B\rho = P/Z$, where $\rho$ is the gyroradius of the particle, and $P$ is its total momentum, measured in \GeVc}.
Apart from the TPCs which are placed downstream of the target, beam counter scintillator A and S1, a telescope of three Beam Position Detectors (BPDs), and veto scintillator V1 are placed before the target.
The A and the S1 detectors are square-shaped plastic scintillators ($6{\times}6$\,\cmsq) equipped with 2 and 4 fast PMTs, respectively.
The two scintillators are separated by approximately 236\,m.
Different isotopes in the beam are identified based on the time-of-flight (\tof) measured between the A and the S1 detectors.
The S1 detector is placed approximately 36\,m upstream of the target and is additionally used to determine the charge of the incoming beam particle.
The BPDs are multiwire proportional chambers with an Ar/CO$_2$ (85/15) gas mixture, and are used to monitor the beam profile.
Each BPD measures the position of the triggered beam particle in the $x$-$y$ plane, which is then used to reconstruct the beam particle track.
The veto scintillator V1 is placed in front of BPD-3 and is an annular disc shaped detector with a diameter of $10$\,\cm and a 1\,\cm central hole.
It is used in the trigger logic to detect and reject highly divergent particles incident on the target.
These upstream detectors are used to define the trigger logic for the beam particle identification.

\subsection{Pilot Run for Fragmentation Studies}
\label{sec:pilotRun}

To study the feasibility of performing nuclear fragmentation measurements at SPS energies, a pilot run was conducted in December 2018.
Its objective was to examine the production of light secondary nuclear fragments, like lithium (Li), beryllium (Be), and boron (B) through \iso{12}{C}+p interactions at $p>10$~\AGeVc.
This was achieved by directing a high-intensity beam consisting of \iso{208}{Pb} nuclei at 13.5~\AGeVc, extracted from the SPS onto the primary target on the H2 beam line, known as T2.
The T2 target is a 160\,mm-long beryllium plate serving as the primary target.
Nuclear fragments resulting from this \iso{208}{Pb}+Be interaction are then transported to the NA61/SHINE experiment situated in EHN1 of the CERN North Area, about 600\,m downstream of the T2 target, via a set of two mass spectrometers.
To study the fragmentation of \iso{12}{C}, the beam-line spectrometer magnets were tuned to select $A/Z=2$ nuclei.
A spread in the longitudinal momentum component, $p_z$ of the fragments is attributed to the Fermi motion.
This effect combined with the momentum acceptance of the beam-line $\Delta p_z/p_z$ (${\approx}1\%$), allows neighboring isotopes with $A/Z<2$ and $A/Z>2$ to be transported to the \NASixtyOne facility.

\begin{figure}
\centering
\includegraphics[height=0.35\textwidth,angle=-90]{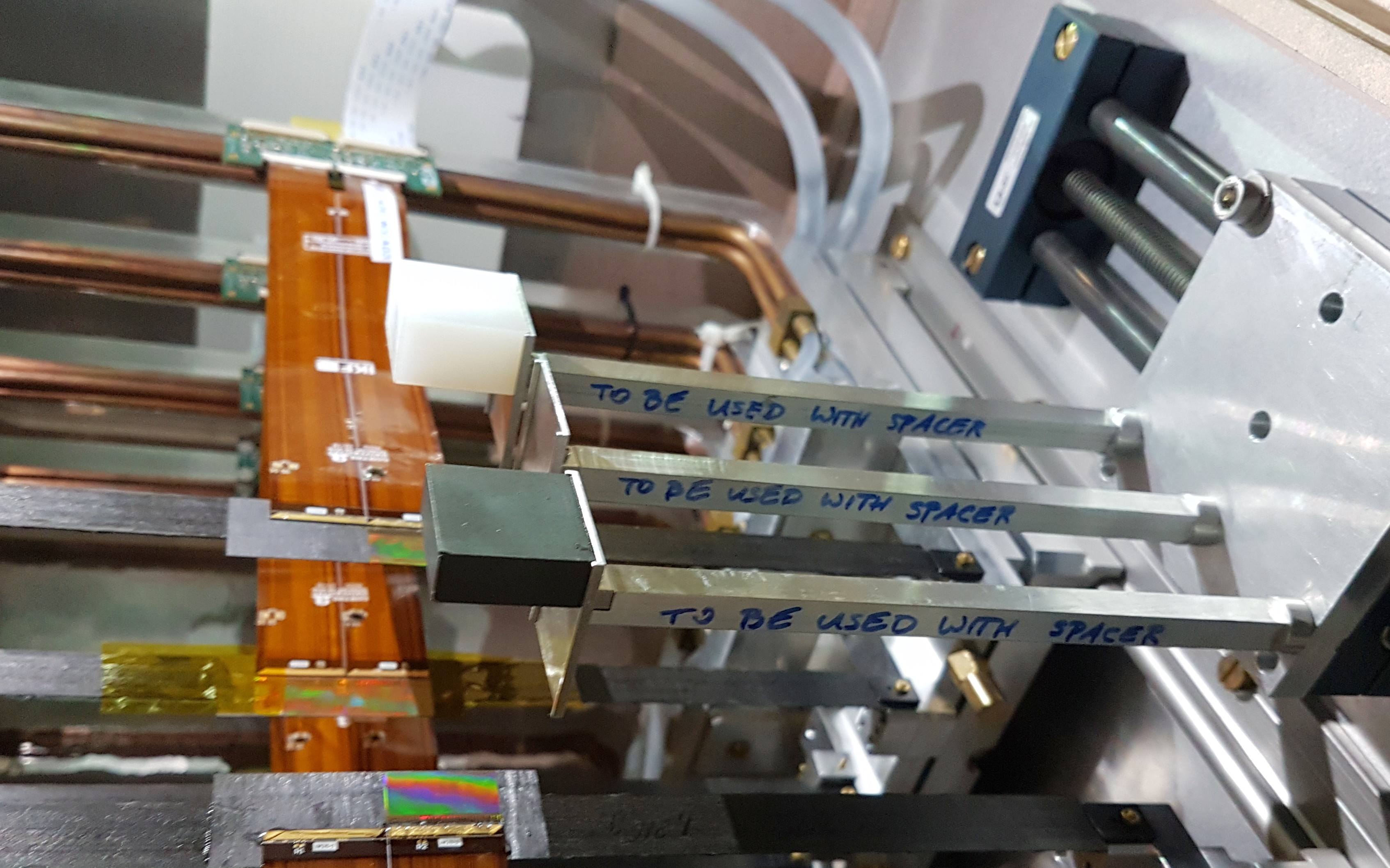}
\caption{Target holder assembly with the two targets namely graphite (C, \emph{left}) and polyethylene (PE, \emph{right}).
The assembly is mounted on a servo-controlled rail operated remotely to switch between the target settings.
The empty target holder seen in the middle was used to record OUT data (see text for details).}
\label{fig:target}
\end{figure}

The rigidity $R$ of a nucleus consisting of $A$ nucleons and charge $Z$ is defined as $R=p_\mathrm{A}(A/Z)$ where $p_A$ is the momentum per nucleon measured using the units \AGeVc.
The current parameterization models of the nuclear cross sections show no dependence on the momentum of the fragmenting projectile nucleus beyond $p_\mathrm{A}>10$\,\GeVc.
Therefore, the rigidity and hence the momentum value was selected such as to be within the operating range delivered by the SPS, and simultaneously to achieve the physics goal of this measurement.
The rigidity of the beam for the pilot run was set to 27\,\GVc, resulting in a momentum per nucleon $p_\mathrm{A}=13.5$\,\GeVc.

Since, we are interested in studying fragmentation of \iso{12}{C} on a proton target, a polyethylene (hereafter abbreviated as PE) block of dimensions $L{\times}B{\times}H = 2.5{\times}2.5{\times}1.5$\,cm$^3$ was used as the primary target material.
Additionally, a graphite (hereafter abbreviated as C) block of dimensions $L{\times}B{\times}H = 2.5{\times}2.5{\times}1.0$\,cm$^3$ was used to account for \iso{12}{C}+C interactions and subtract it from the PE measurements.
Finally, to measure the interactions of the beam outside the target, e.g.\ in the beam counters upstream of the target and the TPC support structures downstream of the target, approximately 10\% of the total events were recorded with an empty target holder. This setting is denoted as OUT.

The target holder was mechanically moved to alternate between the three target settings, namely PE, C (termed as IN), and OUT during the course of the run (see \cref{fig:target}).
The magnetic field was set to 59\% of the maximum field strength of 9\,T\,m in the superconducting vertex magnets (VTX1 and VTX2).
The optimum was determined by performing simulations of tracking the beam particles through different magnetic field settings.
The goal of this optimization was to guide the fragments into the MTPC by ensuring that they hit a minimal amount of material as possible along the curved trajectory.
A total of $\sci{1.1}{6}$ events were recorded, including all three target settings during the 3-day data-taking period. The next section describes the event selection procedure before and after the beam-target interaction, beginning with the selection of \iso{12}{C} ions as the projectiles, and selection of resultant fragments as measured in the MTPC-L.

\begin{figure}
\centering
\def\h{0.4}
\includegraphics[height=\h\textwidth]{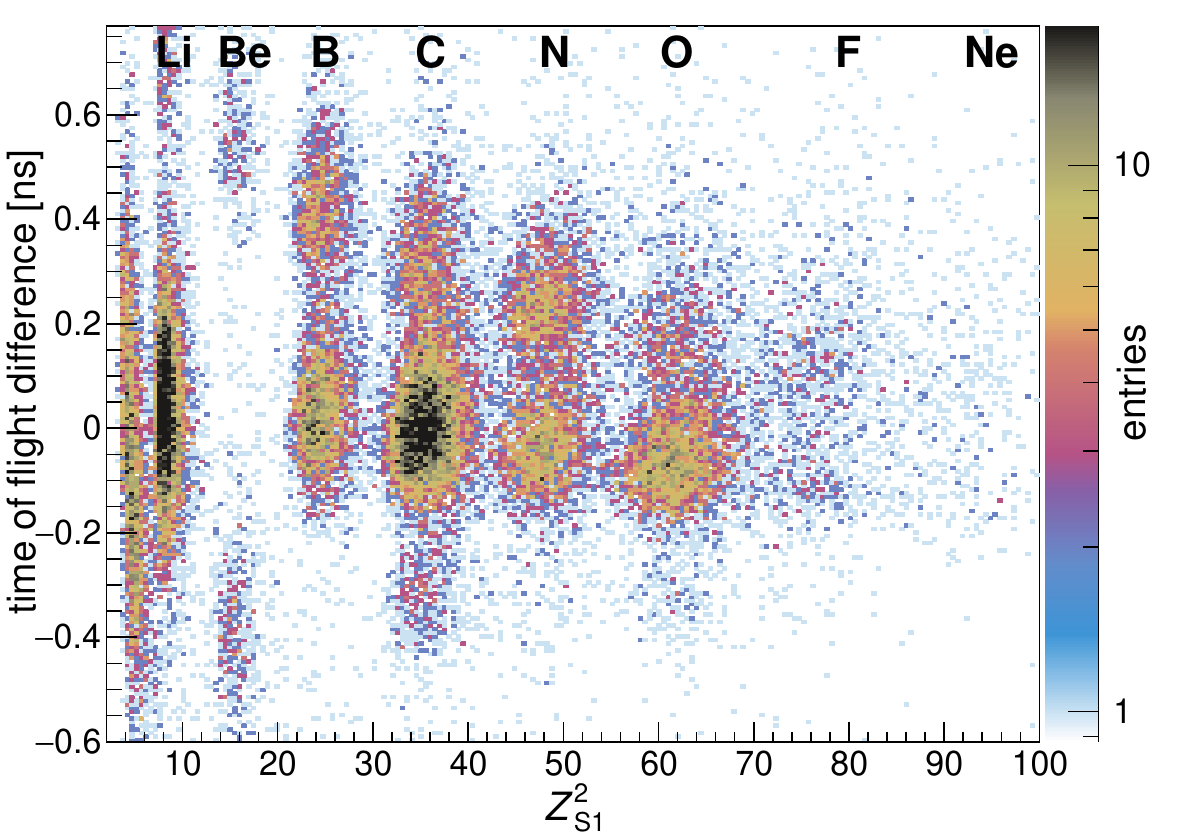}\hfill
\includegraphics[height=\h\textwidth]{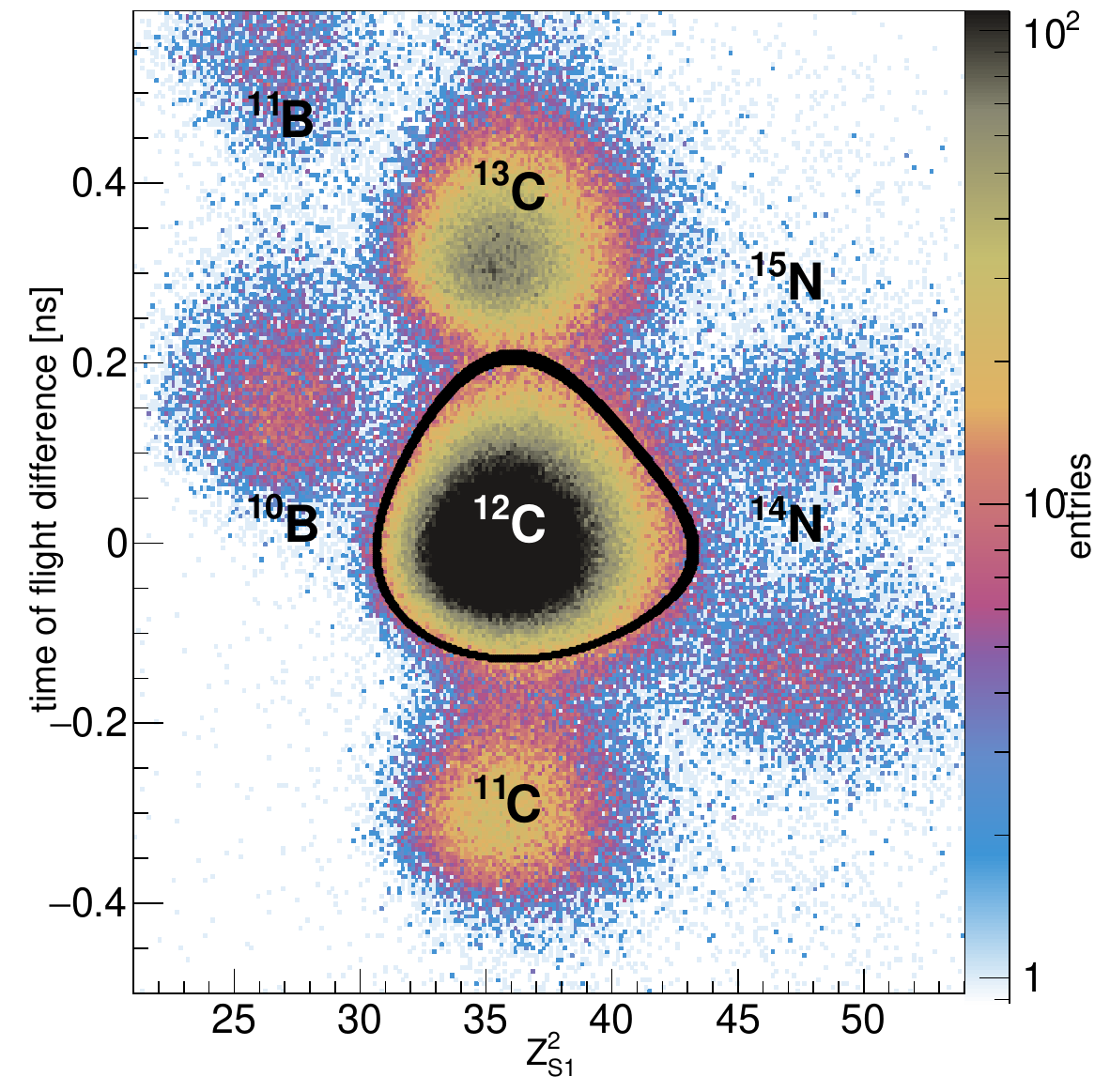}
\caption{The time of flight difference, $\Delta t$ versus the S1 signal, $Z^2_\text{S1}$, distribution of the secondary beam nuclei produced from primary Pb fragmentation, for the beam trigger (left) and carbon trigger (right).
The \dedx in S1 determines the charge represented by the square root of the S1 signal on the $x$-axis and the isotopes separated by the time-of-flight difference between the A and S1 scintillators are shown on the $y$-axis.
The $A/Z=2$ nuclei are at $\Delta t = 0$\,ps. The optimal \iso{12}{C} selection is shown as a black contour in the right plot.}
\label{fig:beamComp}
\end{figure}

\section{Data Selection}
\label{sec:data}
Nuclei deposit energy (\dedx) in the TPC gas mixture proportional to their squared charge, \Zsq, producing ionization electrons.
The electrons then drift in the TPC electric field toward the segmented top plate of the chamber where their position, arrival times, and total number are measured.
The segmented top plate is equipped with electronic readout units called padrows and each padrow is further divided into $1{\times}1$\,\cmsq cells called pads.
The total number of padrows/TPC is proportional to its size and is 90 for MTPCs, 72 for VTPCs, and 7 for the GTPC.
They provide measurement points of the charge clusters produced inside the chamber in the $x$-$z$ plane.
Combining it with the knowledge of the drift of the electrons with a constant velocity along the $y$-direction enables us to reconstruct a 3-dimensional trajectory of the traversing particle.
The reconstructed tracks corresponding to every event are stored in a structured tree format with their 3-dimensional position and charge. The data is then calibrated for further use in the analysis~\cite{Sutter2019}.

In this section, we discuss the event-level and track-level selection of the data used for the analysis. We broadly classify the selection into two categories, namely the upstream selection, which involve identification of the secondary beam-ion \iso{12}{C} as the projectile, based on information from beamline detectors before the target, and selection of downstream events, associated with track selection of the nuclear fragments produced in \iso{12}{C}-target interactions, as measured in the TPCs.

\subsection{Upstream Selection of Secondary Beam Ions}
\label{sec:upSel}

The distribution of secondary beam ions produced by fragmenting primary \iso{208}{Pb} nucleus is shown in \cref{fig:beamComp}.
For the upstream selection of the beam, we operate only with carbon trigger events.
The online trigger to select \iso{12}{C} nuclei in the beam is defined using signals from the two scintillators S1 and V1 as, $(\text{S1} \wedge \overline{\text{V1}})$.
The energy deposit signal in the S1 scintillator was used to determine the charge of the incident beam particle and to trigger on the incoming carbon nuclei.
Two crucial pieces of information to make an offline selection of a particular nucleus as the primary beam, are the \Zsq equivalent signal in S1 for charge selection, and the difference in the recorded arrival times of the beam, $\Delta t$ from the A-S1 system, for identifying a particular isotope.
The time of flight (\tof) of a nucleus with charge $Z$ and mass number $A$, recorded between two scintillators separated by a distance $L$ is given by the formula, $t=L/(\beta c)$, where $\beta=p/E$, $p$ is its longitudinal momentum expressed in terms of the rigidity as $p=RZ$, $E$ is the energy of the nucleus, and $c$ is the speed of light.
The energy is calculated as $E=\sqrt{p^2 + m^2}=\sqrt{(RZ)^2 + (Au)^2}$, where $u$ is the atomic mass unit.
Therefore, the difference $\Delta t$ in the time of flight between two isotopes of mass numbers $A_1$ and $A_2$ is given by,
\begin{equation*}
\Delta t =
  \frac{L}{c}\left(
    \sqrt{1+\left(\frac{A_{1}u}{RZ} \right)^2} - \sqrt{1+\left(\frac{A_{2}u}{RZ} \right)^2}
  \right).
\end{equation*}
It is evident that $\Delta t$ depends on the masses of the two isotopes.
The measured \tof resolution from fitting the 1-dimensional \tof distribution of \iso{12}{C} is $\delta(\Delta t) = 61$\,ps, compared to the measured \tof difference $\Delta t\approx300$\,ps for two neighboring isotopes \iso{13}{C} and \iso{12}{C}.

To estimate the number of recorded events corresponding to various isotopes present in the beam composition, a fit was performed on the 2-dimensional \tof vs.\ \Zsq distribution of the upstream fragments constituting the beam (\cref{fig:beamComp}, \emph{right}).
The model was described with a 2-D Gaussian function with exponential tails, fit in a range $20.0\leqslant Z^2\leqslant54.0$ and $(-0.5\leqslant\Delta t\leqslant0.7)$\,ns (see \cref{app:A4} for further details on the upstream selection of \iso{12}{C}).
The \iso{12}{C} selection cut is shown in \cref{fig:beamComp} as a black contour.
This cut was determined to be the optimum cut for the analysis of the production of carbon, as well as the boron isotopes.

\begin{table}
\caption{A list of upstream selection cuts applied to the calibrated data to select \iso{12}{C} nuclei as the ion beam.}
\label{tab:1}
\begin{center}
\resizebox{\textwidth}{!}{\begin{tabular}{l c c c c c}
\toprule
\multirow{2}{*}{Selection cut} & \multicolumn{3}{c}{Events ($\times10^3$)} & \multirow{2}{*}{Effciency (\%)} & \multirow{2}{*}{Comment}\\
\cmidrule{2-4}
& PE & C & OUT &\\
\midrule
13.5~\AGeVc & 463 & 410 & 106 & 100.0 &--- \\
Carbon Trigger (T3p) & 372 & 331 & 85 & 80.3 & Online $Z^2=36$ triggered events.\\
$t_\mathrm{(A1, A2)}$ & 323 & 286 & 74 & 86.8 & A-det timing information $t_\text{A} > 0$\,ns\\
$a_\mathrm{(A1, A2)}$ & 290 & 258 & 66 & 89.8 & A-det amplitude, $100\leqslant a_\mathrm{A}\leqslant 1000$ \\
Two BPDs & 253 & 225 & 58 & 87.6 & Beam signal present in at least two BPDs.\\
BPD-3 signal & 235 & 210 & 54 & 93.0 & Good BPD-3 measurement\\
WFA cut & 230 & 205 & 53 & 97.6 & Exclude off-time particles.\\
\iso{12}{C} cut & 150 & 135 & 35 & 66.0 & Offline selection of \iso{12}{C} isotope.\\
\bottomrule
\end{tabular}}
\end{center}
\end{table}

In addition to the beam nuclei selection detailed above, we impose further cuts as part of the beam selection criteria on an event-by-event basis.
These cuts are listed in \cref{tab:1} along with the selected number of events for the three data sets, PE, C, and OUT.
The online carbon trigger is tuned to select $Z^2=36$ nuclei based on their energy deposit in the S1 scintillator.
The $t_\text{(A1,A2)}$ and $a_\text{(A1,A2)}$ cuts select events such that the time and signal amplitude as measured by the two PMTs A1 and A2, of the A-detector, are within the specified operational range of the scintillator.
The beam position as measured by BPD-1 is used to calibrate the amplitude of the S1 detector.
In the case where BPD-1 does not measure a signal, the signals from the other two BPDs can be used to make a linear extrapolation to the position of BPD-1, which can then be used for calibration.
This is ensured by the ``Two BPDs'' cut, which only selects events where the beam signal is present in at least two of the three BPDs.
The BPD-3 $x$-$y$ cut is used to filter out events based on the measured $x$-$y$ position of the beam in BPD-3, to make sure that it hits the target.
The aperture diameter for this selection is $\phi=1.6$\,cm.
All events registered out of this region are rejected.
The BPD measurements are further used to reconstruct beam particle tracks by performing a least squares fit to the measured beam position in the $x$-$z$ and the $y$-$z$ planes. Beam particles separation
cut or Waveform Analyzer (WFA) cut is used to prevent counting multiple tracks due to beam particles closely
spaced in time. Particles arriving closer than 2\,$\upmu$s are rejected (track separation in drift
direction $\left|\Delta{y}\right| < 5.0$ cm). The timing information from the S1 scintillator is used for this
cut.

Particles arriving within the event recording time frame of 2\,$\upmu$s are excluded.
Finally, the \iso{12}{C} isotope is selected as the primary beam particle as per the procedure described in \cref{app:A4}.

\subsection{Downstream Selection of Events}
\label{sec:downSel}

The beam-target interaction produces multiple particle tracks in the TPCs associated with every reconstructed event.
Therefore, fragment tracks with the shortest distance to the extrapolated \iso{12}{C} beam in the $x$ and $y$ directions are used for our analysis.
The distance between reconstructed tracks and the beam extrapolation in the $x$-direction is given by $\delx=x_\text{track} - x_\text{beam}$ and similarly in the $y$-direction by $\dely=y_\text{track}-y_\text{beam}$.
The $x-y$ coordinates of the extrapolated beam track are determined by simulating the passage of beam particles traversing the magnetic field inside VTPC-1 and VTPC-2.
Reconstructed tracks satisfying the criteria $\lvert\delx\rvert<30.0$\,cm for MTPC, and $\lvert\delx\rvert<20.0$\,cm for VTPC-2 and GTPC are selected for the analysis.

\begin{figure}
\centering
\includegraphics[width=0.7\textwidth]{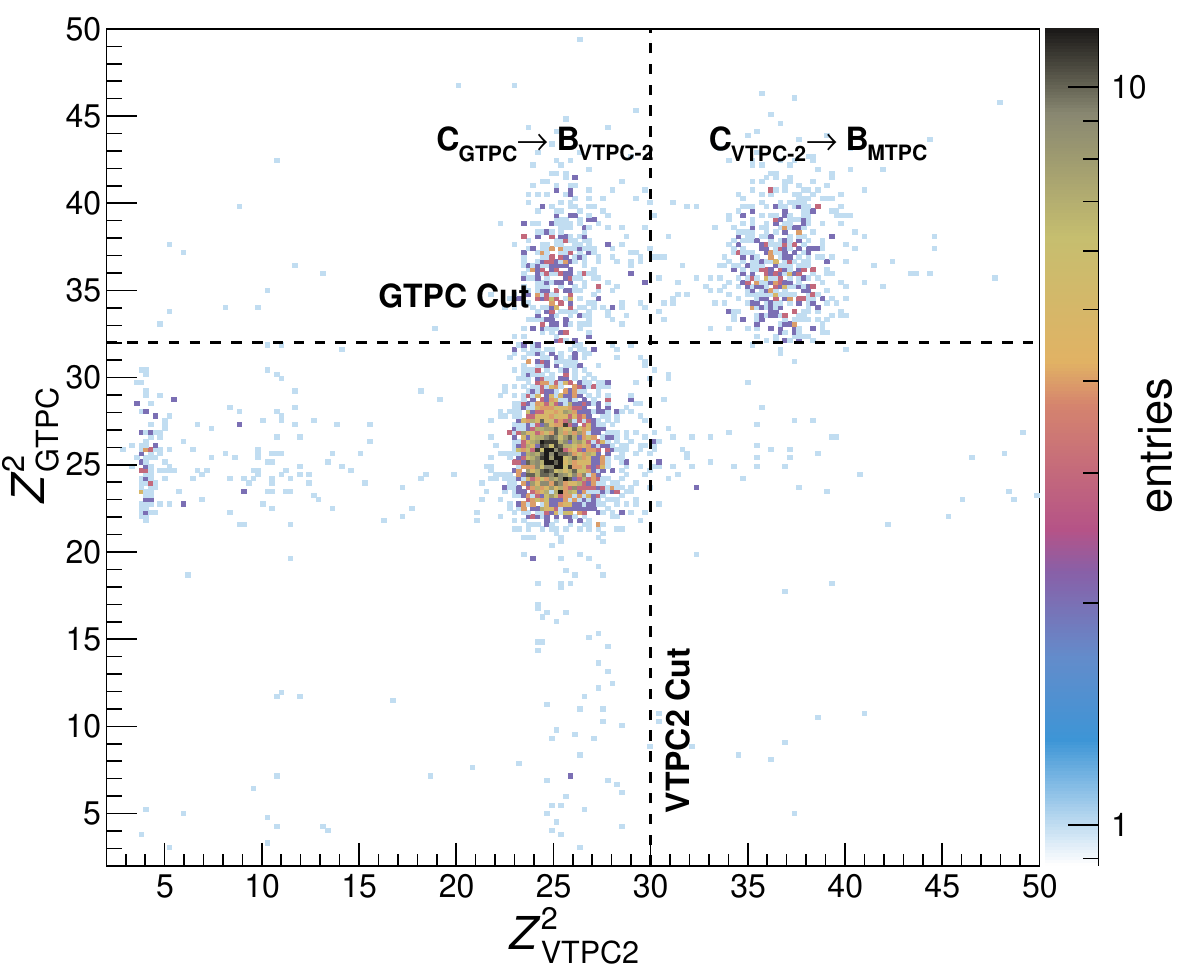}
\caption{Dashed lines represent the re-interaction cuts applied to the calibrated charge signal in GTPC and VTPC-2 to select boron tracks in the MTPC originating from the target. Carbon nuclei fragmenting to boron between GTPC and VTPC-2, and VTPC-2 and MTPC are denoted by C$_\mathrm{GTPC}\to\text{B}_\mathrm{VTPC-2}$ and C$_\mathrm{VTPC-2}\to\text{B}_\mathrm{MTPC}$.}
\label{fig:4}
\end{figure}

Beam particles arriving within a short time window smaller than the readout time lead to pile-up in the detectors, and are called off-time particles.
Such particle tracks are distinguished by measuring the arrival times of the drifting electrons along the $y$-direction of the MTPC.
They are rejected by placing a cut on the nominal beam $y$ position at the end of the MTPC as, $|\Delta y|<5$\,cm.
Moreover, the minimum requirement on the number of charge clusters ($N_\text{clusters}$) produced by a track in the MTPC is 50, and similarly in the VTPC-2 and the GTPC are 15 and 6 respectively.

Carbon fragment tracks in the MTPC are selected by placing cuts on the squared-charge of the track determined from the energy loss \dedx as, $31.0 < Z^2_\text{MTPC} < 44.0$.
Boron fragments tracks are selected using the following cut: $22.5\leqslant Z^2_\text{MTPC}\leqslant27.5$.
Furthermore, as GTPC and VTPC-2 precede the MTPC, we place additional cuts to reject boron isotopes produced from beam fragmentation between these detectors.
We plot \ZsqGTPC vs.\ \ZsqVTPCtwo for tracks corresponding to boron fragments in MTPC-L.
The dominant boron peak at $Z^2 \approx 25$ shows agreement between all the three TPCs.
Nevertheless, two additional peaks at $\ZsqGTPC>33.0$ and $\ZsqVTPCtwo>30.0$ clearly indicate conversion of the \iso{12}{C} beam between GTPC and VTPC-2, and VTPC-2 and MTPC-L, indicated as $\text{C}_\text{GTPC}\to\text{B}_\text{VTPC-2}$ and $\text{C}_\text{VTPC-2}\to\text{B}_\text{MTPC}$ in \cref{fig:4}.
These events are rejected for the boron analysis.

\section{Measurement of Fragments in the MTPC}
\label{sec:frag}

In this section, we present the distribution of the nuclear fragment as measured in the MTPC and explain how the inelastic and boron production cross sections are computed.

\begin{figure}
\centering
\includegraphics[width=\textwidth]{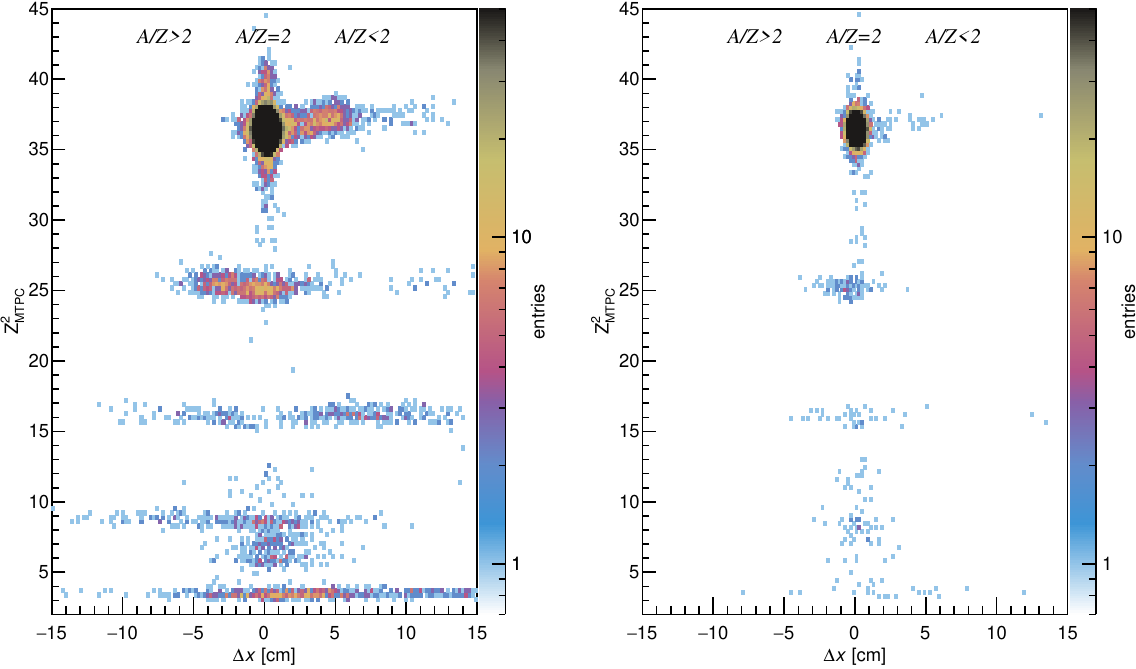}
\caption{\emph{Left:} Secondary fragments produced by the interaction of the \iso{12}{C} beam with the target shown relative to the nominal beam position in the MTPC at $\Delta x_\text{beam}=0.0$\,cm.
The non-interacting beam is seen as the dark blob at this position.
\emph{Right:} Similar distribution of fragments for the OUT case.}
\label{fig:2Dplot}
\end{figure}

A large fraction of the incoming \iso{12}{C} beam particles, incident on the thin target, pass through without any interaction, while a small percentage interacts with the target, producing the lighter nuclei like B, Be, Li, and their respective isotopes.
The charged fragments traversing through the superconducting magnets VTX1 and VTX2 are deflected in the $x$-$y$ plane finally depositing energy in the MTPC.
The nominal position at the end of the MTPC for all $A/Z=2$ nuclei including the \iso{12}{C} beam with respect to the extrapolated beam is at $\Delta x = x_\text{track}- x_\text{beam} = 0.0$\,cm.
Other nuclear fragments are positioned subject to their rigidity, either to the left or the right of this position ($|\Delta x|>0$).
The distribution of such fragments is shown in \cref{fig:2Dplot} for the IN and OUT cases.

To determine the number of tracks corresponding to different isotopes, the distribution is then fitted with a model describing the detector response.
A flat-top Gaussian function together with symmetric exponential tails is used as the detector model to describe the MTPC \delx response.
It aptly combines the momentum acceptance of the beam-line (Rectangular function) convolved with the track resolution of the MTPC (Gaussian distribution).
Each peak corresponds to a type of beam nucleus primary particles and is fitted with this model.
The detector model is mathematically expressed as:
\begin{equation*}
D(\Delta x, \delta x, \sigma_\text{det}, \lambda) =
  \kappa\left(
    \frac{1}{2\delta x}
      \left[
        \erf\left(
          \frac{\Delta x+\frac{\delta x}{2}}{\sigma_\text{det}\sqrt{2}}
        \right) +
        \erf\left(
          \frac{\Delta x-\frac{\delta x}{2}}{\sigma_\text{det}\sqrt{2}}
        \right)
      \right]
    \right) +
  (1-\kappa)\left(
    \frac{1}{2\lambda}
    \exp\left(-\frac{|\Delta x|}{\lambda}\right)
  \right).
\label{eq:detMod}
\end{equation*}
The first term is the flat-top Gaussian function whereas the second term depicts the symmetrical exponential tails characterized by the parameter, $\lambda$, and are attributed to multiple scattering in the target.
Here $\kappa$ denotes the relative fraction of the two mathematical functions in the model.
The fragments produced in the beam-target interaction (secondary particles) are fitted with an additional Gaussian ($G(x,\sigma_\mathrm{F})$) convolved with the detector function ($D(\Delta x, \delta x, \sigma_\text{det}, \lambda)$) to model the spread in the momentum due to their Fermi motion (denoted as $\sigma_\text{F}$ and called the Fermi width hereafter).
The fragment \iso{11}{C} had the highest yield relative to any other isotopes.
Hence its Fermi width, $\sigma_\text{F}(^{11}\text{C})$ was used as an internal reference value in the fit to compute the corresponding widths of low yield fragments (i) in the data, e.g.,\iso{10}{C}, and the boron isotopes.
Therefore, we set $\sigma_\text{F}(i) = \alpha \sigma_\text{F}(^{11}\text{C})$ in the fit.
The scaling factor $\alpha$ is determined by a Monte Carlo method simulating the transport of the fragment nucleus to the MTPC through the $\vec{B}$-field in the VTPCs.
In principle, the method randomizes the momentum 3-vector, $\vec{p}\equiv(p_x, p_y, p_z)$ for each nucleon of the fragment nucleus in its rest frame quantifying the Fermi motion, and then performs a Lorentz boost to the lab frame to get the final values of the 3 components of $\vec{p}$ at the $z$-position corresponding to the end of the MTPC ($z=730$\,cm).
The distribution of the lateral component of the momentum $p_x$ along with the longitudinal component $p_z$ determines the corresponding Fermi width, $\sigma_\text{F}$.
Since $\sigma_\text{F}$ is a function of the mass number of the fragment, $\sigma_\text{F}(^{11}\text{C}) = \sigma_\text{F}(^{11}\text{B})$ is set and, similarly, $\sigma_\text{F}(^{10}\text{C}) = \sigma_\text{F}(^{10}\text{B})$.
The scaling factor $\alpha$ for each of the isotopes in the fit is given in \cref{tab:2}.
\begin{figure}
\centering
\includegraphics[width=\textwidth]{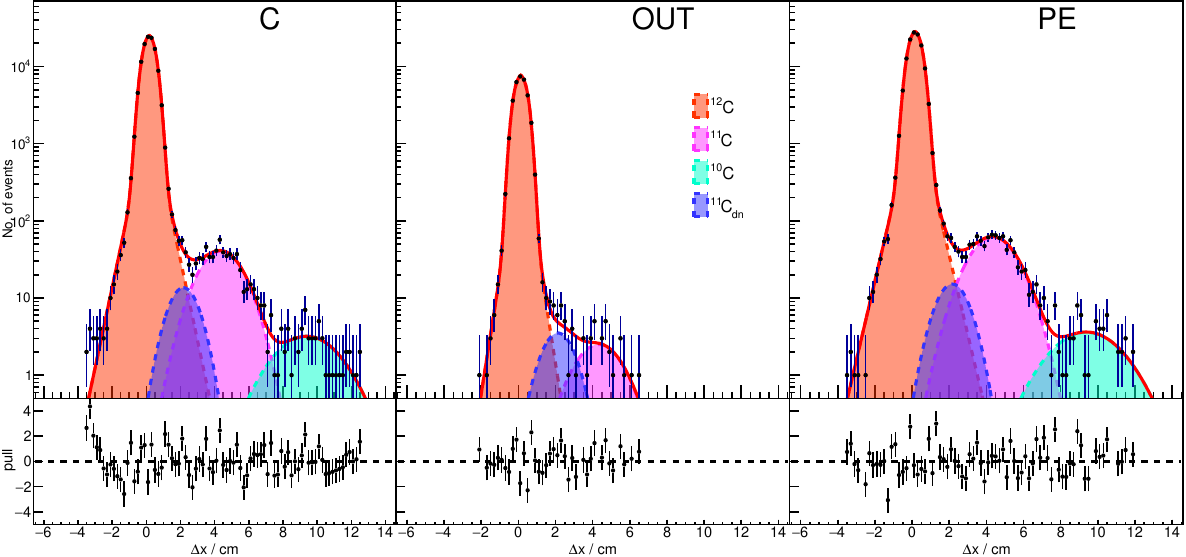}
\caption{Results from the combined fit performed to the \delx-distribution of carbon fragments in the MTPC-L for the C, OUT and PE target settings.
The orange peak is the \iso{12}{C} beam, while the pink and teal peaks correspond to the \iso{11}{C} and \iso{10}{C} fragments respectively.
The \iso{11}{C} fragments produced from the interaction of the beam in the detector support structure is denoted as \iso{11}{C}$_\text{dn}$ and is shown in blue. The black markers show data, $d$, and the solid red curve shows the fit model $m$. Fit residuals are shown in the lower panel, called the pull and is defined as, $\frac{(d - m)}{\sqrt{m}}$.}
\label{fitC}
\end{figure}
\begin{figure}
\centering
\includegraphics[width=\textwidth]{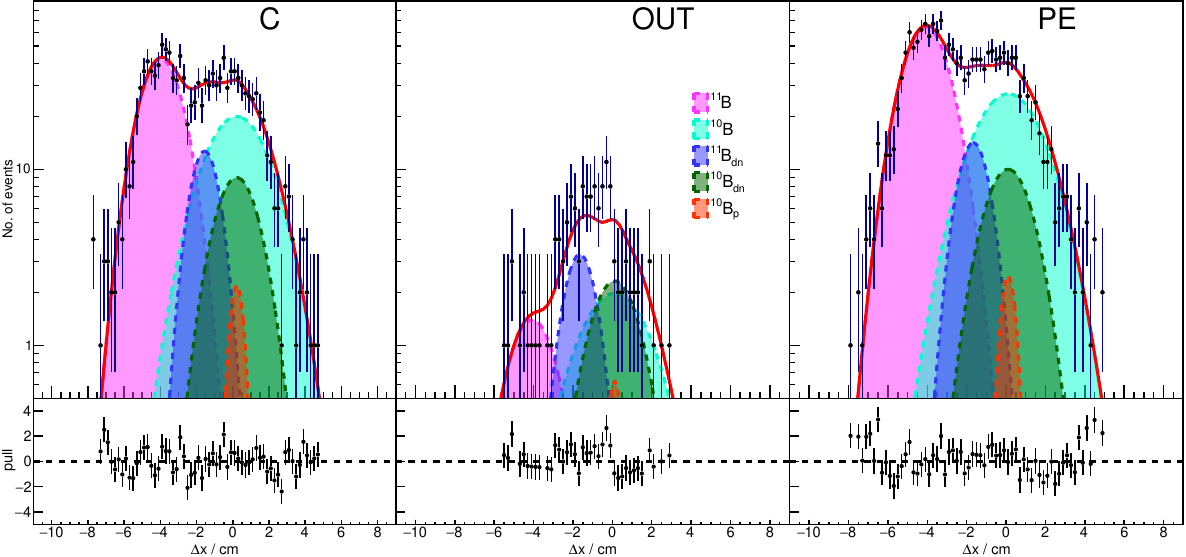}
\caption{Same as \cref{fitC} but for boron fragments.
Various peaks corresponding to the boron isotopes produced in the target (\iso{11}{B} (pink), \iso{10}{B} (teal)), and in the detectors (\iso{11}{B}$_\text{dn}$ (blue) and \iso{10}{B}$_\text{dn}$ (green)) is shown.
The orange peak depicts the primary \iso{10}{B} impurity nuclei as a result of upstream selection of \iso{12}{C}.}
\label{fitB}
\end{figure}
\begin{table}
\caption{The fit parameters for the carbon and boron fragments $\Delta x$-distribution in the MTPC.
Parameters in square brackets are fixed.
$^{\dagger}$\,The experimental parameters are shared amongst the three data sets while the fragment parameter, $\sigma_\text{F}$ and $\alpha$ are given for each isotope.
The parameter $\alpha$ is used for scaling the fragment width with respect to the \iso{11}{C} width as, $\sigma_\text{F} = \alpha\sigma_\text{$^{11}$C}$ (see text for details).
The superscript `p' stands for primary and corresponds to the nucleus present in the beam composition, whereas the superscript `dn' corresponds to fragments produced inside the magnetic field, downstream of the target.}
\label{tab:2}
\begin{center}
\resizebox{\textwidth}{!}{
\begin{tabular}{l l c c c c c c c}
\toprule
\multirow{2}{*}{Element} & \multirow{2}{*}{Isotope} & \multicolumn{5}{c}{Experimental parameters$^{\dagger}$} & \multicolumn{2}{c}{Fragment parameter} \\
\cline{3-7}\cline{8-9} \\
& & $\delta x$ (cm) & $\sigma_\text{det}$ (cm) & $\Theta$ (GV\,cm/$c$) & $\lambda$ (cm) & $\kappa$ & $\alpha$ & $\sigma_\text{F}$ (cm) \\
\midrule
\multirow{4}{*}{Carbon - } & 12$^\text{p}$ & \multirow{4}{*}{$0.416\pm0.006$} & \multirow{4}{*}{$0.252\pm0.005$}  & \multirow{4}{*}{$1244\pm22$} & \multirow{4}{*}{$0.42\pm0.02$} & \multirow{4}{*}{$0.83\pm0.02$} & - & [0.0]\\
   & 11 & & & & & & 1.00 & $1.17\pm0.05$\\
       & 10 & & & & & & 1.52 & $1.77\pm0.08$\\
       & 11$^\text{dn}$ & & & & & & 0.69 & $0.80\pm0.03$\\
\midrule
\multirow{5}{*}{Boron -} & $10^\text{p}$ &\multirow{5}{*}{[0.416]} & \multirow{5}{*}{[0.252]} & \multirow{5}{*}{[1244]} & \multirow{5}{*}{[0.42]} & \multirow{5}{*}{[0.83]}& - & [0.00]\\
 & 10 &  &  &   & & & 1.52 & [1.77]\\
 & 11 & & & & & & 1.00 & [1.17]\\
       & 11$^\text{dn}$ & & & & & & 0.69 & [0.80] \\
       & 10$^\text{dn}$ & & & & & & 0.67 & [0.78] \\
\bottomrule
\end{tabular}
}
\end{center}
\end{table}

The complete model used for the fit is given by,
\begin{equation*}
D(\Delta x, \delta x, \sigma_\text{det}, \lambda) \otimes G(\Delta x,\sigma_\text{F}) =
  D(\Delta x, \delta x, \sigma_\text{det}, \lambda) \otimes
  \frac{1}{\sigma_\text{F}\sqrt{2\pi}} \exp\left(-\frac{\Delta x^2}{2\sigma_\text{F}^{2}}\right).
\label{eq:model}
\end{equation*}
The convolution is computed over the entire fitting range, $\abs{x} < 22.0$\,cm.
A combined log-likelihood fit of the three datasets namely PE, C, and OUT is performed on the $\Delta x$-distribution of carbon fragments using the MINUIT minimization procedure~\cite{James:1975dr}.
The fit result is shown in \cref{fitC}, and the best-fit parameter values retrieved from the fit are given in \cref{tab:2}.

The production probability of the fragments produced downstream of the target, inside the magnetic field (denoted with a subscript `dn' in \cref{fitC,fitB}), is independent of the target setting.
Therefore, the normalization parameter of these peaks is shared among the data-sets in the combined fit.
The beam-line acceptance is calculated from $\delta x/x_\text{beam}$, where $x_\text{beam}$ corresponds to the $x$-coordinate of the beam after deflecting through the vertex magnetic field.
From the best-fit $\delta x$ value, we get $\delta x/x_\text{beam} = 0.42/46.98 \approx 0.9\%$, which is consistent with the experimentally set value ${\approx}1\%$.
The $x$-position corresponding to a particular nuclear fragment is related to its rigidity as, $x\propto 1/R$.
Given that the $x$ position of the beam ($x_\text{beam}$) and its rigidity ($R_\text{beam} = 27\,$\GVc) is known, the relative positions of the neighboring isotopes along the $x$ direction can be calculated using their respective rigidities ($R_\text{iso}$).
We introduce a parameter, $\Theta$ which identifies with the bending power of the magnetic field.
Hence the $x$ positions of the isotopes relative to the \iso{12}{C} beam are characterized by this single parameter, as:
\begin{equation*}
\Delta x_\text{iso} =
  (x_\text{iso} - x_\text{beam}) =
  \Theta  \left(\frac{1}{R_\text{iso}} - \frac{1}{R_\text{beam}} \right).
\end{equation*}
Furthermore, the energy loss of the beam inside the target reduces its total momentum and induces an additional constant shift in $x_\text{beam}$.
This is calculated based on the change in the total momentum of the beam as, $\Delta p_\text{beam} = (p_A A_\text{beam} - p_{\dedx})$, where $p_{\dedx}$ quantifies the energy loss.
Therefore, the beam rigidity can be written as, $R_\text{beam} = \Delta p_\text{beam}/Z$, to further calculate the relative positions of the isotopes as discussed earlier.
For empty target holder (OUT), we set $p_{\dedx} = 0.0$\,\GeVc, while the same for targets retrieved from the fit is $p_{\dedx} = 0.12$\,\GeVc for PE, and $p_{\dedx} \approx 0.14$\,\GeVc.
These values are within 10\% of energy loss calculated from Bethe-Bloch formula for a minimum ionizing particle.

The experimental parameters ($\delta x, \sigma_\text{det}, \Theta, \lambda$) for the boron fit were fixed to those determined in the carbon fit depicted by the square brackets in \cref{tab:2}.
The uncertainties on these parameters were propagated in the boron fits to study the systematic effect on the final measured cross sections.
We found that the results were altered by ${<}1\%$, which is negligible compared to the statistical uncertainty of the measurement (see \cref{sec:res}).

\begin{table}
    \caption{The number of recorded beam events $N_\text{b}$ and fragment tracks corresponding to the three datasets, PE, C, and OUT, as measured in the MTPC-L.
    In the case \iso{12}{C}, the probability computed here is the survival probability, whereas for other nuclei, it corresponds to its production probability.}
    \label{tab:stats}
    \begin{center}
    \resizebox{\textwidth}{!}{\begin{tabular}{l c  c c  c c  c c c c }
        \toprule
        Dataset & $N_\text{beam}$ & $N_{^{12}\text{C}}$ & $P_{^{12}\text{C}\to^{12}\text{C}}$ & $N_{^{11}\text{C}}$ & $P_{^{12}\text{C}\to^{11}\text{C}}$ & $N_{^{11}\text{B}}$ & $P_{^{12}\text{C}\to^{11}\text{B}}$ & $N_{^{10}\text{B}}$ & $P_{^{12}\text{C}\to^{10}\text{B}}$\\
        \midrule
        PE & 150883 & 128270 & $0.8501\pm0.0009$ & 908 & $0.0060\pm0.0002$ & 894 & $0.0059\pm0.0002$ & 565 & $0.0037\pm0.0002$\\
        C & 135284 & 115506 & $0.8538\pm0.0010$ & 596 & $0.0044\pm0.0002$ & 586 & $0.0043\pm0.0002$ & 421 & $0.0031\pm0.0001$\\
        OUT & 34990 & 32108 & $0.9176\pm0.0015$ & 37 & $0.0011\pm0.0002$ & 29 & $0.0008\pm0.0002$ & 41 & $0.0012\pm0.0002$\\
        \bottomrule
    \end{tabular}}
    \end{center}
\end{table}

\section{Analysis}
\label{sec:analysis}

\begin{figure}[t]
\includegraphics[width=\textwidth]{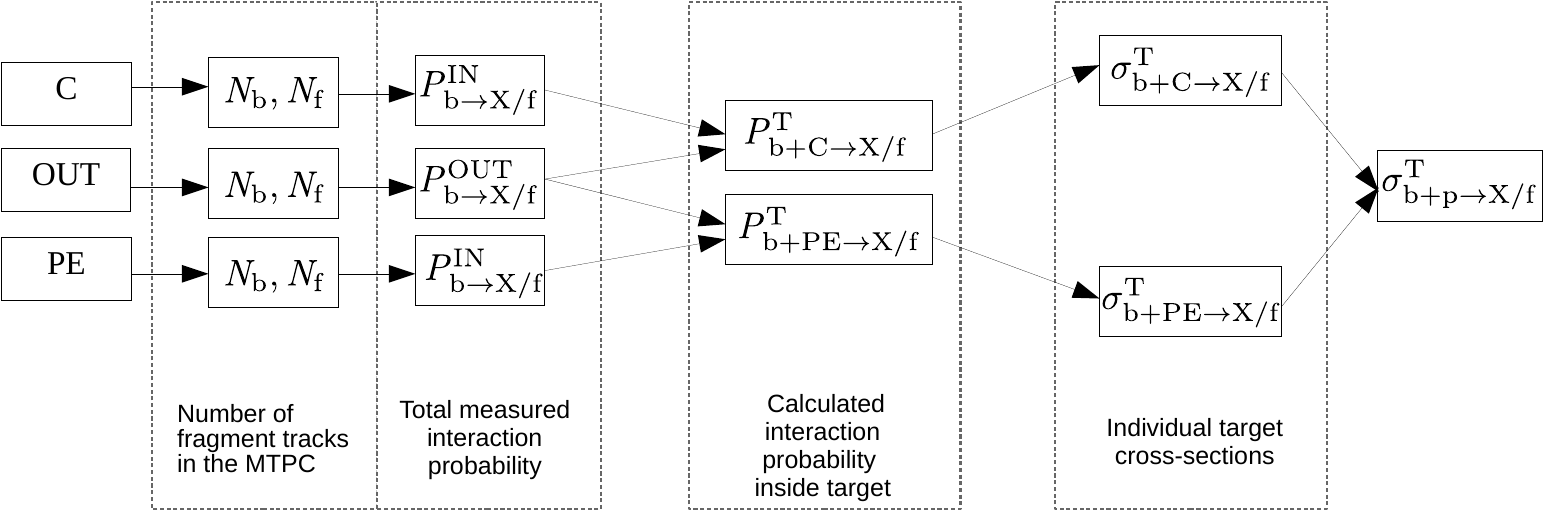}
\caption{The flow of the analysis for cross section calculation on a proton target derived from the cross sections on PE and C targets. The mass-changing and charge-changing inelastic reactions are represented as b$\to$X whereas the isotope production cross sections are written as b$\to$f (see~\cref{sec:ccXsec,sec:isoXsec} for further details).}
\label{fig:analysis}
\end{figure}

The primary aim of our analysis is to determine inelastic cross sections leading to the production of the isotopes \iso{11}{C}, \iso{10}{B} and \iso{11}{B} nuclei, from \iso{12}{C}+p interactions.
We also present a measurement of the total \iso{12}{C} mass-changing, and the charge-changing cross sections.
The flow chart of the analysis is shown in \cref{fig:analysis}. The analysis framework is mainly based on the formalism described in Ref.\cite{Bennemann2023_1000164132}.
Let us denote the parent nucleus as \b (beam), and the resulting fragment nucleus of interest as \f.
Then, the probability that the nucleus \b passes without undergoing an interaction, retaining its identity, is written as \Ps{b}{b} for the beam nucleus and the same for the fragment nucleus as \Ps{f}{f}.
In the following, we call this term the survival probability of the corresponding nucleus.

The experimental facility is broadly divided into four zones which are identified as interaction regions for the beam.
These zones are the upstream region (`up') comprising the beam-line counters and scintillators (e.g.\ S1, BPD-1/2/3 etc.), before the target, the second the target itself (T), the third a region close to but downstream of the target denoted as `VD' and corresponds to the enclosed He-filled Vertex Detector chamber.
The fourth region includes the TPCs and their support structures downstream of the target, denoted as `dn'.
The total survival probability of nucleus \b with the target (IN) is a directly measurable quantity as determined from the fit normalization (see \cref{tab:stats}). It is expressed as the ratio of the measured number of \b tracks in the MTPC $N_\text{b}$, to the total number of beam particles incident on the target $N_\text{beam}$, as $\PP{IN}{b}{b} = N_\text{b}/N_\text{beam}$.
To obtain an analytical formula of the in-target interaction probability of the fragmenting nucleus, the total survival probability can further be expressed as a product of the four survival probabilities corresponding to the interaction zones described above.
Hence, the total survival probability of the nucleus \b for a given target setting T is:
\begin{equation}
  \PP{IN}{b}{b} = \PP{up}{b}{b} \, \PP{T}{b}{b} \, \PP{VD}{b}{b} \, \PP{dn}{b}{b}.
\label{eq:1}
\end{equation}
For the empty target holder case (OUT), the target survival probability of \b $\left(\PP{T}{b}{b}\right)$ inside the target is equal to 1.
Therefore, the total survival probability in this case is written as:
\begin{equation}
  \PP{OUT}{b}{b} = \PP{up}{b}{b} \, \PP{VD}{b}{b} \, \PP{dn}{b}{b}.
\label{eq:2}
\end{equation}
The upstream and downstream probabilities in the equations above are independent of the target setting and are therefore the same in the IN and OUT cases.
Furthermore, dividing \cref{eq:1} by \cref{eq:2} gives the total probability that the nucleus \b does not undergo any inelastic interaction inside the target.
Since probabilities are conserved quantities, we can obtain the total mass-changing interaction probability of \b inside the target, $\PP{T}{b}{X}$, which can be written as
\begin{equation}
  \PP{T}{b}{X} = \left(1 - \PP{T}{b}{b}\right) = \left(1 - \frac{\PP{IN}{b}{b}}{\PP{OUT}{b}{b}}\right).
\label{eq:3}
\end{equation}
Here the symbol `X' denotes any nuclear fragment other than \b indicating that it has fragmented due to a mass-changing reaction.
Similarly, the mass-changing probability and the cross section value can be determined for any nucleus of interest by using \cref{eq:3}.
This generality enables us to make auxiliary measurements of various nuclei crucial for our calculations, as will be made more clear in the following sections (see appendix~\cref{app:A1} for further details on auxiliary measurements).
In addition, nucleus \b fragmenting to \f inside the target, where ($A_\text{b}-A_\text{f}=1$), which then inelastically interacts inside the target itself, will appear as a mass-changing reaction of \b.
This is a two step reaction happening inside a thin target, and, hence, the product of the two interaction probabilities is $\approx 10^{-6}$ (see~\cref{app:A5}), significantly smaller compared to the survival probabilities in \cref{eq:1}.

The analysis for the charge-changing interaction is derived in the same way as described above, to obtain a final expression for the in-target interaction probability, such as~\cref{eq:3} (see~\cref{sec:ccXsec} for further details). Similarly, the final expression for the in-target production of a fragment \f, is derived in~\cref{sec:isoXsec}.

The expression for in-target interaction as given in \cref{eq:3} is a set of two equations, each corresponding to the two different targets (T) namely, PE and C.
The interaction cross section, $\sigma^\text{T}$ is related to the interaction probability $P^\text{T}$ as,
\begin{equation}
P^\text{T} = 1 - \exp\left(-d_\text{T}/\lambda\right)
\label{eq:11}
\end{equation}
where, $d_\text{T}$ is the thickness of the target, $\lambda = 1/(n_\text{T}\,\sigma^\text{T})$ is the interaction length, and $n_\text{T}$ is the number density of the target.
It is expressed in terms of the target density, $\rho_\text{T}$, target molar mass, $M_\text{T}$ and Avogadro's constant $N_\text{A}$ as $n_\text{T}=N_\text{A}\,\rho_\text{T}/M_\text{T}$.
Hence, the cross section can be determined by making these substitutions and re-arranging the terms in \cref{eq:11} to give the final expression as,
\begin{equation}
\sigma^\text{T} =
  -\frac{M_\text{T}}{N_\text{A}\,\rho_\text{T}\,d_\text{T}}\ln\left(1- P^\text{T}\right).
\label{eq:12}
\end{equation}
Given that the primary objective of our study is to measure the \iso{12}{C}+p charge-changing, mass-changing, and production cross sections of lighter isotopes, the final cross section is computed as:
\begin{equation}
\sigma^\text{p}_{^{12}\text{C}\to\text{X/f}} =
  \frac{1}{2}\left(
    \sigma^\text{PE}_{^{12}\text{C}\to\text{X/f}} - \sigma^\text{C}_{^{12}\text{C}\to\text{X/f}}
  \right)
\label{eq:13}
\end{equation}
Here X denotes the charge/mass-changing reaction of \iso{12}{C} and \f denotes the production of specific isotope (in our case \iso{11}{C}, \iso{11}{B}, and \iso{10}{B}).
The factor of 1/2 arises from the fact that the ratio \(\text{C}:\text{H} = 1:2\) in each unit of the polyethylene molecule must be considered when subtracting the \iso{12}{C}+C contribution from the \iso{12}{C}+PE interactions.

The final cross section values are subject to further corrections, taking into account the purity of the beam, the selection of the fragments in the MTPC produced in the beam-target interaction, inelastic interaction of fragments inside the target and uncertainty in the target density. All the corrections calculated in our analysis are data-driven and are $\mathcal{O}$(10\%), whereas the corresponding systematic uncertainties are $\mathcal{O}$(1-2\%) of our measurement. The corrections and systematic uncertainties are tabulated in~\cref{tab:corr}, and are detailed in~\cref{app:B}.

\section{Results}
\label{sec:res}
\begin{figure}[t]
\centering
	\includegraphics[width=0.8\textwidth]{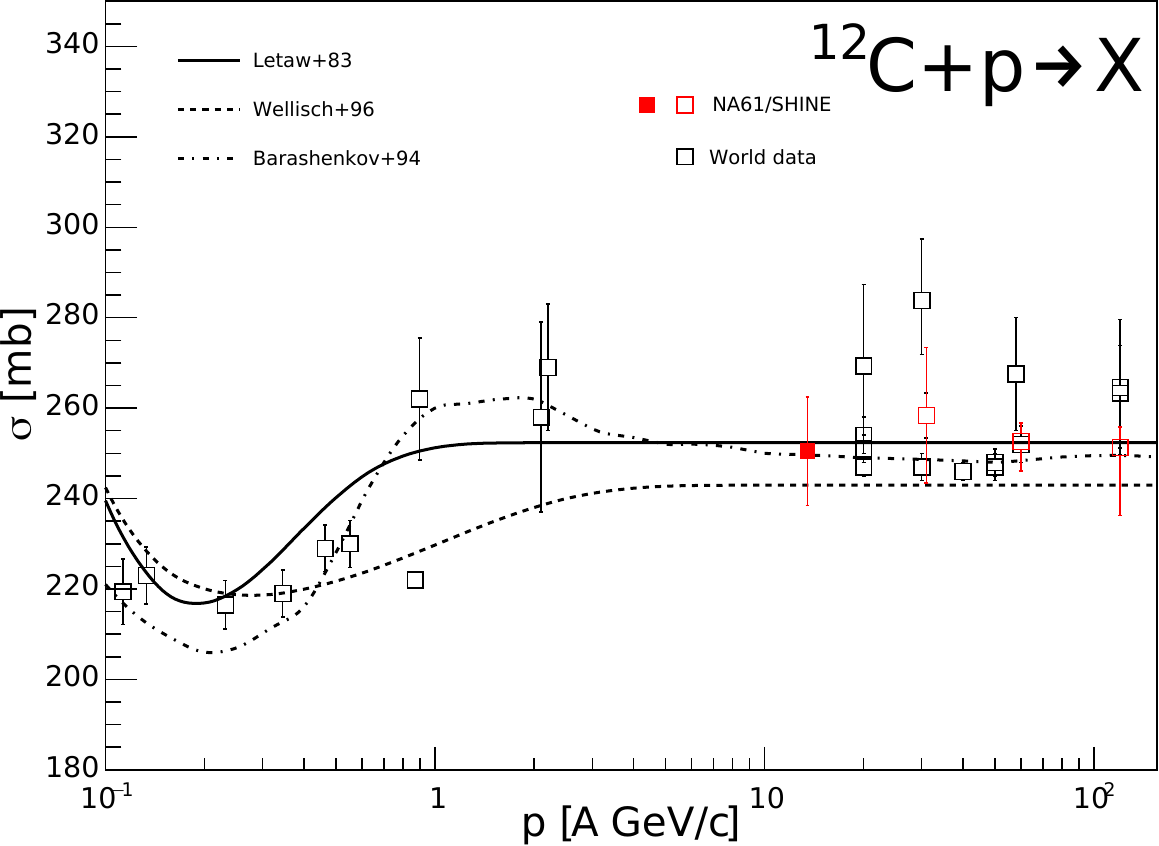}
	\caption{The \iso{12}{C}+p mass-changing cross section computed from this analysis (solid red square) along with previously reported values (open black squares). Previous measurements by NA61/SHINE in p+C interactions are shown as open red squares~\cite{NA61:2016,NA61:2019}.
 The lines represent various momentum-dependent parameterizations of the mass-changing cross section~\cite{Barashenkov94,Letaw1983,Wellisch:1996xm}.}
	\label{fig:resultMCX}
\end{figure}

The total reaction cross section along with the charge-changing cross
section of the C nucleus, and the proton-nucleus inelastic interaction
cross section has been previously measured by many
groups \cite{Bellettini:1966zz,Denisov:1973zv,Carroll:1978hc,Mahajan:2013awa}
including \NASixtyOne, using a proton beam at various momenta.  The
data show a momentum dependence of the cross section at energies
$E<2$\,GeV, while at higher momenta the behavior is asymptotic.
Previous \NASixtyOne measurements of the inelastic cross sections
using a proton beam on a graphite target were performed at 31~\AGeVc,
60~\AGeVc, and 120~\AGeVc beam momenta.  The \iso{12}{C}+p
mass-changing cross section are computed using \cref{eq:13}
(\cref{fig:resultMCX}).  The overlaying lines correspond to the
function describing the energy dependence of the cross section at low
and high energies as detailed in
Refs.~\cite{Barashenkov94,Letaw1983,Wellisch:1996xm}.  Our result is
in good agreement with previous studies and the model lines. While the
systematic uncertainties arising due to various corrections are
calculated as described in \cref{app:B}, the total estimated errors on
our current measurements are dominated by statistical uncertainty on
the ${\approx}10\%$ level. The analysis of the charge-changing cross
sections for \iso{12}{C} was validated in an independent
analysis~\cite{Marta2024}.  The corrections applied to the calculated
cross sections and the systematic uncertainty are given
in \cref{tab:corr}.

\begin{figure}[t]
    \centering \includegraphics[width=\textwidth]{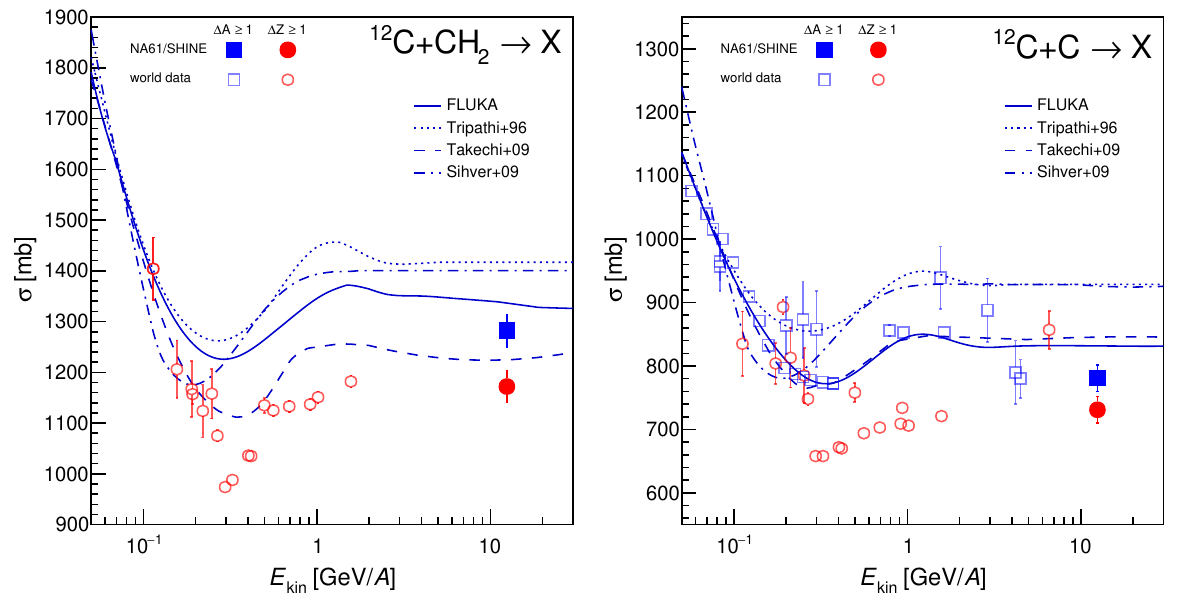} \caption[cross section]{The
   charge-changing (${\Delta{Z}\geq1}$) and mass-changing
   (${\Delta{A}\geq1}$) cross section of \iso{12}{C} on CH$_2$ (left)
   and C (right) targets as function of kinetic energy per nucleon.
   The results from this work are shown as solid circles and squares.
   Previous measurements of the charge-changing cross sections are
   shown as open circles and previous measurements of the
   mass-changing or inelastic cross section are displayed as open
   squares~\cite{Ferrando1988,Golovchenko,Webber:1990_PRC,SCHALL1996221,CHULKOV2000330,Jaros:1978,Akisenko:1980,Zhang:2002,Zheng:2002,Fang:2000,Yan:2020,Kanungo:2016}.
   The lines are adapted from Ref.~\cite{SIHVER2012812} for the
   parameterizations of the inelastic cross section of
   Refs.~\cite{TRIPATHI1996347,Takechi2009,Sihver2009} and the
   predictions of the FLUKA2024.1
   model~\cite{Ferrari:2005zk} were obtained
   from Ref.~\cite{flukapriv}.}  \label{fig:ChCh}
\end{figure}
The charge- and mass-changing cross section on C and
PE target are compared to model predictions and previous data in
Fig.~\ref{fig:ChCh} as a function of kinetic energy per nucleon
$E_\text{kin}=(\sqrt{p^2 + m^2} - m)/A$, where $m$ denotes the mass of
the nucleus\footnote{Note that the model predictions as well as some
of the data in Fig.~\ref{fig:ChCh} are for the inelastic cross
section, whereas we measured the mass-changing cross section. However,
the inelastic cross section is almost identical to the mass-changing
cross section at high energy (few mb difference above 0.5 GeV
according to the FLUKA model).}.  As can be seen, our measurement
provides important new constraints at high beam momenta, where the
cross sections are predicted to flatten out.

The highest-energy measurements of the charge-changing cross section
of carbon were derived from data of the AMS detector in
Ref.~\cite{Yan:2020}. The paper reports a numerical value of $857\pm15\,
(\text{stat}.)\pm 26 \, (\text{syst.})$ at 7.5~\AGeVc and states that the
measured cross-section is approximately collision-momentum independent
above 4~\AGeVc. This value of the charge-changing cross section is $+(126 \pm 37)$~mb
higher than our measurement at 13.5~\AGeVc.

\begin{figure}
	\centering
	\includegraphics[width=0.65\linewidth]{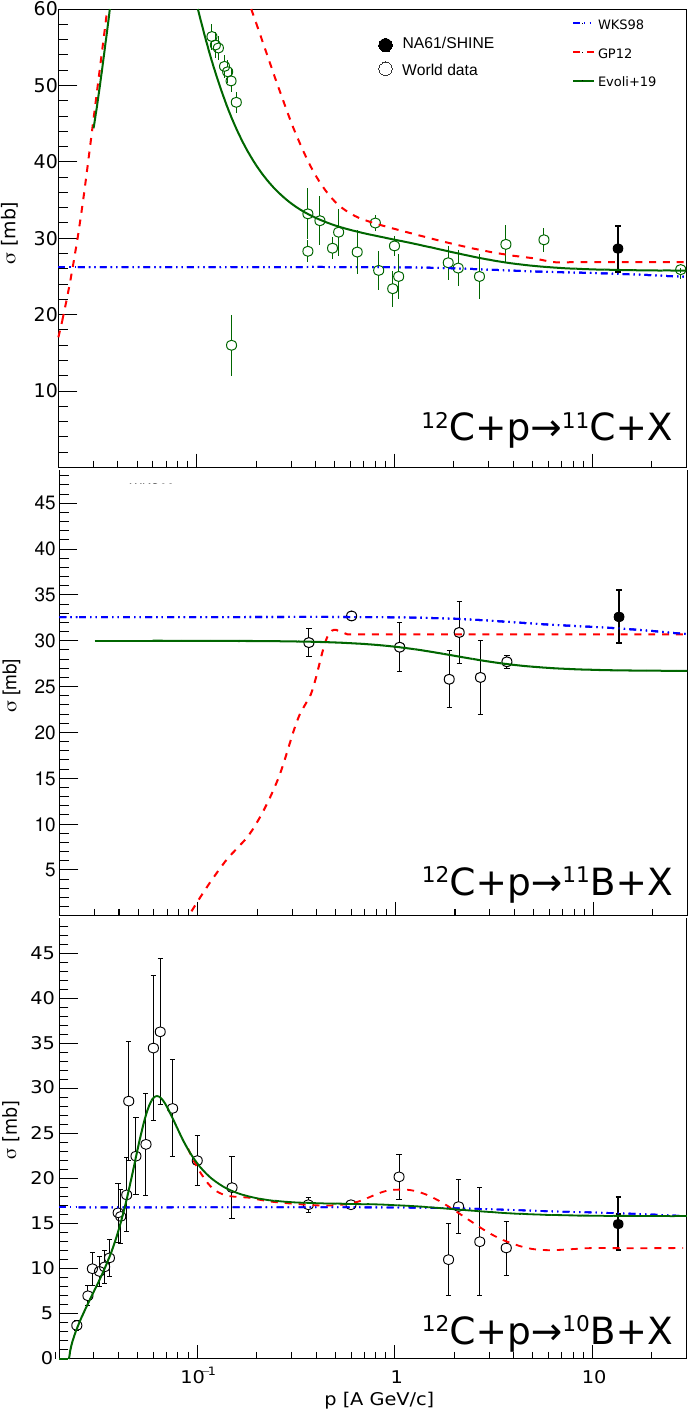}
	\caption{The isotope production cross section of \iso{11}{C} (top), \iso{11}{B} (center), and \iso{10}{B} fragment (bottom) as measured in this work in \iso{12}{C}+p reaction, compared to previous measurements~\cite{Fontes:1977qq,Korejwo:1999,Korejwo:2002,Olson:1983,Webber:1990aipc}.
 The lines represent the parameterization of the cross section with respect to momentum per nucleon.}
	\label{fig:bresults}
\end{figure}

The isotope production cross section results from our analysis are
shown in \cref{fig:bresults}, for the \iso{11}{C}, \iso{11}{B},
and \iso{10}{B} fragments produced in \iso{12}{C}+p interactions.  The
lines represent the parameterization of the cross section as a
function of momentum per nucleon and correspond to the models
GALPROP12, abbreviated as GP12, developed in
Ref.~\cite{Webber:1990pr}, the WKS98 by the authors of
Refs.~\cite{Webber:1998ex,Webber:1998c}, and the Evoli+19 fit, as
given in Ref.~\cite{Evoli:2019pr}.  Our pilot measurement is in good
agreement with the previous high-energy isotopic production cross
section for \iso{11}{C} at $p = 28\AGeVc$~\cite{Cummings:1962pr} and
300\,\AGeVc~\cite{Kaufman:1976xh} (outside the range of
Fig.~\ref{fig:bresults}) which reported a cross section of
$25.9\pm1.2$~mb and $24.6\pm1.6$~mb, respectively.

The cumulative production cross section of boron isotopes including
contributions from the decay of \iso{11}{C} and \iso{10}{C} has been
measured previously by many groups. The measurements at $p=25$~\AGeVc reported
in Ref.~\cite{Fontes:1977qq}  are ($59\pm12$)\,mb for the cumulative production of \iso{11}{B} and ($20\pm3$)\,mb for \iso{10}{B}, which is in good
agreement with the cumulative cross sections that can be obtained from
the data presented here at 13.5~\AGeVc.

\begin{table}[t]
	\caption{Corrected \iso{12}{C} mass-changing, charge-changing, and carbon and boron isotope production cross sections calculated in this work.
 The format of the cross section results is $\upsigma^\text{T} \pm \updelta\upsigma^\text{stat.} \pm \updelta\upsigma^\text{syst.}$, where T=PE,C,p}
	\label{tab:5}
	\begin{center}
	\resizebox{\textwidth}{!}{
		\begin{tabular}{
				l
				S[table-format=3.1(3)]
				S[table-format=3.1(3)]
				S[table-format=3.1(3)]
				c
			}
			\toprule
			Reaction & {$\upsigma^\text{PE}$ (mb)} & {$\upsigma^\text{C}$ (mb)} & {$\upsigma^\text{p} = \frac{1}{2}\left(\upsigma^\text{PE} - \upsigma^\text{C}\right)$ (mb)} & Remark \\
			\midrule
			\iso{12}{C}$\to$X & {\SI[parse-numbers=false]{1282\pm 32 \pm 1}{}} & {\SI[parse-numbers=false]{781\pm21\pm5}{}} & {\SI[parse-numbers=false]{250 \pm 12 \pm 2}{}} & Mass-changing, $A_\text{X}<12$ \\
			\iso{12}{C}$\to$X & {\SI[parse-numbers=false]{1172\pm 32\pm 1}{}} & {\SI[parse-numbers=false]{731\pm21\pm5}{}} & {\SI[parse-numbers=false]{221 \pm 12 \pm 2}{}} & Charge-changing, $Z_\text{X}<6$ \\
			\iso{12}{C}$\to$\iso{11}{C} & {\SI[parse-numbers=false]{101 \pm 5\pm1}{}} & {\SI[parse-numbers=false]{43 \pm 3\pm 1}{}} & {\SI[parse-numbers=false]{29 \pm 3 \pm 0}{}} & \hspace{-1.5em}\rdelim\}{3}{*}[\hspace{0.5em}Isotope Production] \\
			\iso{12}{C}$\to$\iso{11}{B} & {\SI[parse-numbers=false]{110 \pm 5\pm 2}{}} & {\SI[parse-numbers=false]{45\pm 3\pm 1}{}} & {\SI[parse-numbers=false]{33 \pm 3 \pm 1}{}} & \\
			\iso{12}{C}$\to$\iso{10}{B} & {\SI[parse-numbers=false]{57\pm 5\pm 1}{}} & {\SI[parse-numbers=false]{26 \pm 3\pm 1}{}} & {\SI[parse-numbers=false]{15 \pm 3 \pm 0}{}} & \\
			\bottomrule
		\end{tabular}
	}
 \end{center}
\end{table}

\section{Summary and Outlook}
\label{sec:con}

In this work, we have presented the nuclear fragmentation cross section measurements from the \NASixtyOne data recorded during a pilot run in 2018. We studied \iso{12}{C}+p interactions using the $\text{CH}_2$ and C targets resulting in the production of lighter nuclear fragments of B, Be, and Li isotopes.
The main aim of this paper is to highlight the isotope production cross section values which are critical inputs in determining the propagation characteristics of cosmic rays in the Galaxy.
The measurements reported in this work demonstrate that conducting nuclear fragmentation studies of intermediate-mass nuclei at \NASixtyOne is feasible.
Our results, although from a pilot run, are in good agreement with previously reported cross section values.
Nevertheless, it is currently dominated by statistical uncertainty.
Recently, \NASixtyOne detector systems have undergone a major upgrade during the Long Shutdown 2 period from 2018 to 2021.
The front-end readout electronic boards of the TPCs were replaced with faster boards significantly boosting the data acquisition rate from 100\,Hz to approximately 1\,kHz.
This implies an almost 10 times more gain in the number of recorded events, and equivalent gain in the precision of our measurements currently limited by low statistics.
A dedicated high statistics run to study fragmentation of various nuclei like C, N, O, and Si is scheduled for the end of 2024.

\section*{Acknowledgments}
We would like to thank the CERN EP, BE, HSE and EN Departments for the
strong support of \NASixtyOne.
This work was supported by
the Hungarian Scientific Research Fund (grant NKFIH 138136\slash137812\slash138152 and TKP2021-NKTA-64),
the Polish Ministry of Science and Higher Education
(DIR\slash WK\slash\-2016\slash 2017\slash\-10-1, WUT ID-UB), the National Science Centre Poland (grants
2014\slash 14\slash E\slash ST2\slash 00018, %
2016\slash 21\slash D\slash ST2\slash 01983, %
2017\slash 25\slash N\slash ST2\slash 02575, %
2018\slash 29\slash N\slash ST2\slash 02595, %
2018\slash 30\slash A\slash ST2\slash 00226, %
2018\slash 31\slash G\slash ST2\slash 03910, %
2020\slash 39\slash O\slash ST2\slash 00277), %
the Norwegian Financial Mechanism 2014--2021 (grant 2019\slash 34\slash H\slash ST2\slash 00585),
the Polish Minister of Education and Science (contract No. 2021\slash WK\slash 10),
the European Union's Horizon 2020 research and innovation programme under grant agreement No. 871072,
the Ministry of Education, Culture, Sports,
Science and Tech\-no\-lo\-gy, Japan, Grant-in-Aid for Sci\-en\-ti\-fic
Research (grants 18071005, 19034011, 19740162, 20740160 and 20039012,22H04943),
the German Research Foundation DFG (grants GA\,1480\slash8-1 and project 426579465),
the Bulgarian Ministry of Education and Science within the National
Roadmap for Research Infrastructures 2020--2027, contract No. D01-374/18.12.2020,
Serbian Ministry of Science, Technological Development and Innovation (grant
OI171002), Swiss Nationalfonds Foundation (grant 200020\-117913/1),
ETH Research Grant TH-01\,07-3, National Science Foundation grant
PHY-2013228 and the Fermi National Accelerator Laboratory (Fermilab),
a U.S. Department of Energy, Office of Science, HEP User Facility
managed by Fermi Research Alliance, LLC (FRA), acting under Contract
No. DE-AC02-07CH11359 and the IN2P3-CNRS (France).\\

The data used in this paper were collected before February 2022.

\newpage
\bibliographystyle{na61Utphys}
{\footnotesize\raggedright
\bibliography{fragmentation}
}

\newpage
{\Large The \NASixtyOne Collaboration}
\bigskip
\begin{sloppypar}

\noindent
{H.\;Adhikary\,\href{https://orcid.org/0000-0002-5746-1268}{\includegraphics[height=1.7ex]{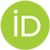}}\textsuperscript{\,11}},
{P.\;Adrich\,\href{https://orcid.org/0000-0002-7019-5451}{\includegraphics[height=1.7ex]{orcidlogo.png}}\textsuperscript{\,13}},
{K.K.\;Allison\,\href{https://orcid.org/0000-0002-3494-9383}{\includegraphics[height=1.7ex]{orcidlogo.png}}\textsuperscript{\,24}},
{N.\;Amin\,\href{https://orcid.org/0009-0004-7572-3817}{\includegraphics[height=1.7ex]{orcidlogo.png}}\textsuperscript{\,4}},
{E.V.\;Andronov\,\href{https://orcid.org/0000-0003-0437-9292}{\includegraphics[height=1.7ex]{orcidlogo.png}}\textsuperscript{\,20}},
{I.-C.\;Arsene\,\href{https://orcid.org/0000-0003-2316-9565}{\includegraphics[height=1.7ex]{orcidlogo.png}}\textsuperscript{\,10}},
{M.\;Bajda\,\href{https://orcid.org/0009-0005-8859-1099}{\includegraphics[height=1.7ex]{orcidlogo.png}}\textsuperscript{\,14}},
{Y.\;Balkova\,\href{https://orcid.org/0000-0002-6957-573X}{\includegraphics[height=1.7ex]{orcidlogo.png}}\textsuperscript{\,16}},
{D.\;Battaglia\,\href{https://orcid.org/0000-0002-5283-0992}{\includegraphics[height=1.7ex]{orcidlogo.png}}\textsuperscript{\,23}},
{A.\;Bazgir\,\href{https://orcid.org/0000-0003-0358-0576}{\includegraphics[height=1.7ex]{orcidlogo.png}}\textsuperscript{\,11}},
{J.\;Bennemann\,\href{https://orcid.org/0009-0007-2300-3799}{\includegraphics[height=1.7ex]{orcidlogo.png}}\textsuperscript{\,4}},
{S.\;Bhosale\,\href{https://orcid.org/0000-0001-5709-4747}{\includegraphics[height=1.7ex]{orcidlogo.png}}\textsuperscript{\,12}},
{M.\;Bielewicz\,\href{https://orcid.org/0000-0001-8267-4874}{\includegraphics[height=1.7ex]{orcidlogo.png}}\textsuperscript{\,13}},
{A.\;Blondel\,\href{https://orcid.org/0000-0002-1597-8859}{\includegraphics[height=1.7ex]{orcidlogo.png}}\textsuperscript{\,3}},
{M.\;Bogomilov\,\href{https://orcid.org/0000-0001-7738-2041}{\includegraphics[height=1.7ex]{orcidlogo.png}}\textsuperscript{\,2}},
{Y.\;Bondar\,\href{https://orcid.org/0000-0003-2773-9668}{\includegraphics[height=1.7ex]{orcidlogo.png}}\textsuperscript{\,11}},
{A.\;Bravar\,\href{https://orcid.org/0000-0002-1134-1527}{\includegraphics[height=1.7ex]{orcidlogo.png}}\textsuperscript{\,27}},
{W.\;Bryli\'nski\,\href{https://orcid.org/0000-0002-3457-6601}{\includegraphics[height=1.7ex]{orcidlogo.png}}\textsuperscript{\,19}},
{J.\;Brzychczyk\,\href{https://orcid.org/0000-0001-5320-6748}{\includegraphics[height=1.7ex]{orcidlogo.png}}\textsuperscript{\,14}},
{M.\;Buryakov\,\href{https://orcid.org/0009-0008-2394-4967}{\includegraphics[height=1.7ex]{orcidlogo.png}}\textsuperscript{\,20}},
{A.F.\;Camino\textsuperscript{\,26}},
{M.\;\'Cirkovi\'c\,\href{https://orcid.org/0000-0002-4420-9688}{\includegraphics[height=1.7ex]{orcidlogo.png}}\textsuperscript{\,21}},
{M.\;Csan\'ad\,\href{https://orcid.org/0000-0002-3154-6925}{\includegraphics[height=1.7ex]{orcidlogo.png}}\textsuperscript{\,6}},
{J.\;Cybowska\,\href{https://orcid.org/0000-0003-2568-3664}{\includegraphics[height=1.7ex]{orcidlogo.png}}\textsuperscript{\,19}},
{T.\;Czopowicz\,\href{https://orcid.org/0000-0003-1908-2977}{\includegraphics[height=1.7ex]{orcidlogo.png}}\textsuperscript{\,11}},
{C.\;Dalmazzone\,\href{https://orcid.org/0000-0001-6945-5845}{\includegraphics[height=1.7ex]{orcidlogo.png}}\textsuperscript{\,3}},
{N.\;Davis\,\href{https://orcid.org/0000-0003-3047-6854}{\includegraphics[height=1.7ex]{orcidlogo.png}}\textsuperscript{\,12}},
{A.\;Dmitriev\,\href{https://orcid.org/0000-0001-7853-0173}{\includegraphics[height=1.7ex]{orcidlogo.png}}\textsuperscript{\,20}},
{P.~von\;Doetinchem\,\href{https://orcid.org/0000-0002-7801-3376}{\includegraphics[height=1.7ex]{orcidlogo.png}}\textsuperscript{\,25}},
{W.\;Dominik\,\href{https://orcid.org/0000-0001-7444-9239}{\includegraphics[height=1.7ex]{orcidlogo.png}}\textsuperscript{\,17}},
{J.\;Dumarchez\,\href{https://orcid.org/0000-0002-9243-4425}{\includegraphics[height=1.7ex]{orcidlogo.png}}\textsuperscript{\,3}},
{R.\;Engel\,\href{https://orcid.org/0000-0003-2924-8889}{\includegraphics[height=1.7ex]{orcidlogo.png}}\textsuperscript{\,4}},
{G.A.\;Feofilov\,\href{https://orcid.org/0000-0003-3700-8623}{\includegraphics[height=1.7ex]{orcidlogo.png}}\textsuperscript{\,20}},
{L.\;Fields\,\href{https://orcid.org/0000-0001-8281-3686}{\includegraphics[height=1.7ex]{orcidlogo.png}}\textsuperscript{\,23}},
{Z.\;Fodor\,\href{https://orcid.org/0000-0003-2519-5687}{\includegraphics[height=1.7ex]{orcidlogo.png}}\textsuperscript{\,5,18}},
{M.\;Friend\,\href{https://orcid.org/0000-0003-4660-4670}{\includegraphics[height=1.7ex]{orcidlogo.png}}\textsuperscript{\,7}},
{M.\;Ga\'zdzicki\,\href{https://orcid.org/0000-0002-6114-8223}{\includegraphics[height=1.7ex]{orcidlogo.png}}\textsuperscript{\,11}},
{K.E.\;Gollwitzer\textsuperscript{\,22}},
{O.\;Golosov\,\href{https://orcid.org/0000-0001-6562-2925}{\includegraphics[height=1.7ex]{orcidlogo.png}}\textsuperscript{\,20}},
{V.\;Golovatyuk\,\href{https://orcid.org/0009-0006-5201-0990}{\includegraphics[height=1.7ex]{orcidlogo.png}}\textsuperscript{\,20}},
{M.\;Golubeva\,\href{https://orcid.org/0009-0003-4756-2449}{\includegraphics[height=1.7ex]{orcidlogo.png}}\textsuperscript{\,20}},
{K.\;Grebieszkow\,\href{https://orcid.org/0000-0002-6754-9554}{\includegraphics[height=1.7ex]{orcidlogo.png}}\textsuperscript{\,19}},
{F.\;Guber\,\href{https://orcid.org/0000-0001-8790-3218}{\includegraphics[height=1.7ex]{orcidlogo.png}}\textsuperscript{\,20}},
{P.G.\;Hurh\,\href{https://orcid.org/0000-0002-9024-5399}{\includegraphics[height=1.7ex]{orcidlogo.png}}\textsuperscript{\,22}},
{S.\;Ilieva\,\href{https://orcid.org/0000-0001-9204-2563}{\includegraphics[height=1.7ex]{orcidlogo.png}}\textsuperscript{\,2}},
{A.\;Ivashkin\,\href{https://orcid.org/0000-0003-4595-5866}{\includegraphics[height=1.7ex]{orcidlogo.png}}\textsuperscript{\,20}},
{A.\;Izvestnyy\,\href{https://orcid.org/0009-0009-1305-7309}{\includegraphics[height=1.7ex]{orcidlogo.png}}\textsuperscript{\,20}},
{N.\;Karpushkin\,\href{https://orcid.org/0000-0001-5513-9331}{\includegraphics[height=1.7ex]{orcidlogo.png}}\textsuperscript{\,20}},
{M.\;Kie{\l}bowicz\,\href{https://orcid.org/0000-0002-4403-9201}{\includegraphics[height=1.7ex]{orcidlogo.png}}\textsuperscript{\,12}},
{V.A.\;Kireyeu\,\href{https://orcid.org/0000-0002-5630-9264}{\includegraphics[height=1.7ex]{orcidlogo.png}}\textsuperscript{\,20}},
{R.\;Kolesnikov\,\href{https://orcid.org/0009-0006-4224-1058}{\includegraphics[height=1.7ex]{orcidlogo.png}}\textsuperscript{\,20}},
{D.\;Kolev\,\href{https://orcid.org/0000-0002-9203-4739}{\includegraphics[height=1.7ex]{orcidlogo.png}}\textsuperscript{\,2}},
{Y.\;Koshio\,\href{https://orcid.org/0000-0003-0437-8505}{\includegraphics[height=1.7ex]{orcidlogo.png}}\textsuperscript{\,8}},
{S.\;Kowalski\,\href{https://orcid.org/0000-0001-9888-4008}{\includegraphics[height=1.7ex]{orcidlogo.png}}\textsuperscript{\,16}},
{B.\;Koz{\l}owski\,\href{https://orcid.org/0000-0001-8442-2320}{\includegraphics[height=1.7ex]{orcidlogo.png}}\textsuperscript{\,19}},
{A.\;Krasnoperov\,\href{https://orcid.org/0000-0002-1425-2861}{\includegraphics[height=1.7ex]{orcidlogo.png}}\textsuperscript{\,20}},
{W.\;Kucewicz\,\href{https://orcid.org/0000-0002-2073-711X}{\includegraphics[height=1.7ex]{orcidlogo.png}}\textsuperscript{\,15}},
{M.\;Kuchowicz\,\href{https://orcid.org/0000-0003-3174-585X}{\includegraphics[height=1.7ex]{orcidlogo.png}}\textsuperscript{\,18}},
{M.\;Kuich\,\href{https://orcid.org/0000-0002-6507-8699}{\includegraphics[height=1.7ex]{orcidlogo.png}}\textsuperscript{\,17}},
{A.\;Kurepin\,\href{https://orcid.org/0000-0002-1851-4136}{\includegraphics[height=1.7ex]{orcidlogo.png}}\textsuperscript{\,20}},
{A.\;L\'aszl\'o\,\href{https://orcid.org/0000-0003-2712-6968}{\includegraphics[height=1.7ex]{orcidlogo.png}}\textsuperscript{\,5}},
{M.\;Lewicki\,\href{https://orcid.org/0000-0002-8972-3066}{\includegraphics[height=1.7ex]{orcidlogo.png}}\textsuperscript{\,12}},
{G.\;Lykasov\,\href{https://orcid.org/0000-0002-1544-6959}{\includegraphics[height=1.7ex]{orcidlogo.png}}\textsuperscript{\,20}},
{V.V.\;Lyubushkin\,\href{https://orcid.org/0000-0003-0136-233X}{\includegraphics[height=1.7ex]{orcidlogo.png}}\textsuperscript{\,20}},
{M.\;Ma\'ckowiak-Paw{\l}owska\,\href{https://orcid.org/0000-0003-3954-6329}{\includegraphics[height=1.7ex]{orcidlogo.png}}\textsuperscript{\,19}},
{A.\;Makhnev\,\href{https://orcid.org/0009-0002-9745-1897}{\includegraphics[height=1.7ex]{orcidlogo.png}}\textsuperscript{\,20}},
{B.\;Maksiak\,\href{https://orcid.org/0000-0002-7950-2307}{\includegraphics[height=1.7ex]{orcidlogo.png}}\textsuperscript{\,13}},
{A.I.\;Malakhov\,\href{https://orcid.org/0000-0001-8569-8409}{\includegraphics[height=1.7ex]{orcidlogo.png}}\textsuperscript{\,20}},
{A.\;Marcinek\,\href{https://orcid.org/0000-0001-9922-743X}{\includegraphics[height=1.7ex]{orcidlogo.png}}\textsuperscript{\,12}},
{A.D.\;Marino\,\href{https://orcid.org/0000-0002-1709-538X}{\includegraphics[height=1.7ex]{orcidlogo.png}}\textsuperscript{\,24}},
{H.-J.\;Mathes\,\href{https://orcid.org/0000-0002-0680-040X}{\includegraphics[height=1.7ex]{orcidlogo.png}}\textsuperscript{\,4}},
{T.\;Matulewicz\,\href{https://orcid.org/0000-0003-2098-1216}{\includegraphics[height=1.7ex]{orcidlogo.png}}\textsuperscript{\,17}},
{V.\;Matveev\,\href{https://orcid.org/0000-0002-2745-5908}{\includegraphics[height=1.7ex]{orcidlogo.png}}\textsuperscript{\,20}},
{G.L.\;Melkumov\,\href{https://orcid.org/0009-0004-2074-6755}{\includegraphics[height=1.7ex]{orcidlogo.png}}\textsuperscript{\,20}},
{A.\;Merzlaya\,\href{https://orcid.org/0000-0002-6553-2783}{\includegraphics[height=1.7ex]{orcidlogo.png}}\textsuperscript{\,10}},
{{\L}.\;Mik\,\href{https://orcid.org/0000-0003-2712-6861}{\includegraphics[height=1.7ex]{orcidlogo.png}}\textsuperscript{\,15}},
{S.\;Morozov\,\href{https://orcid.org/0000-0002-6748-7277}{\includegraphics[height=1.7ex]{orcidlogo.png}}\textsuperscript{\,20}},
{Y.\;Nagai\,\href{https://orcid.org/0000-0002-1792-5005}{\includegraphics[height=1.7ex]{orcidlogo.png}}\textsuperscript{\,6}},
{T.\;Nakadaira\,\href{https://orcid.org/0000-0003-4327-7598}{\includegraphics[height=1.7ex]{orcidlogo.png}}\textsuperscript{\,7}},
{M.\;Naskr\k{e}t\,\href{https://orcid.org/0000-0002-5634-6639}{\includegraphics[height=1.7ex]{orcidlogo.png}}\textsuperscript{\,18}},
{S.\;Nishimori\,\href{https://orcid.org/~0000-0002-1820-0938}{\includegraphics[height=1.7ex]{orcidlogo.png}}\textsuperscript{\,7}},
{A.\;Olivier\,\href{https://orcid.org/0000-0003-4261-8303}{\includegraphics[height=1.7ex]{orcidlogo.png}}\textsuperscript{\,23}},
{V.\;Ozvenchuk\,\href{https://orcid.org/0000-0002-7821-7109}{\includegraphics[height=1.7ex]{orcidlogo.png}}\textsuperscript{\,12}},
{O.\;Panova\,\href{https://orcid.org/0000-0001-5039-7788}{\includegraphics[height=1.7ex]{orcidlogo.png}}\textsuperscript{\,11}},
{V.\;Paolone\,\href{https://orcid.org/0000-0003-2162-0957}{\includegraphics[height=1.7ex]{orcidlogo.png}}\textsuperscript{\,26}},
{O.\;Petukhov\,\href{https://orcid.org/0000-0002-8872-8324}{\includegraphics[height=1.7ex]{orcidlogo.png}}\textsuperscript{\,20}},
{I.\;Pidhurskyi\,\href{https://orcid.org/0000-0001-9916-9436}{\includegraphics[height=1.7ex]{orcidlogo.png}}\textsuperscript{\,11}},
{R.\;P{\l}aneta\,\href{https://orcid.org/0000-0001-8007-8577}{\includegraphics[height=1.7ex]{orcidlogo.png}}\textsuperscript{\,14}},
{P.\;Podlaski\,\href{https://orcid.org/0000-0002-0232-9841}{\includegraphics[height=1.7ex]{orcidlogo.png}}\textsuperscript{\,17}},
{B.A.\;Popov\,\href{https://orcid.org/0000-0001-5416-9301}{\includegraphics[height=1.7ex]{orcidlogo.png}}\textsuperscript{\,20,3}},
{B.\;P\'orfy\,\href{https://orcid.org/0000-0001-5724-9737}{\includegraphics[height=1.7ex]{orcidlogo.png}}\textsuperscript{\,5,6}},
{D.S.\;Prokhorova\,\href{https://orcid.org/0000-0003-3726-9196}{\includegraphics[height=1.7ex]{orcidlogo.png}}\textsuperscript{\,20}},
{D.\;Pszczel\,\href{https://orcid.org/0000-0002-4697-6688}{\includegraphics[height=1.7ex]{orcidlogo.png}}\textsuperscript{\,13}},
{S.\;Pu{\l}awski\,\href{https://orcid.org/0000-0003-1982-2787}{\includegraphics[height=1.7ex]{orcidlogo.png}}\textsuperscript{\,16}},
{R.\;Renfordt\,\href{https://orcid.org/0000-0002-5633-104X}{\includegraphics[height=1.7ex]{orcidlogo.png}}\textsuperscript{\,16}},
{L.\;Ren\,\href{https://orcid.org/0000-0003-1709-7673}{\includegraphics[height=1.7ex]{orcidlogo.png}}\textsuperscript{\,24}},
{V.Z.\;Reyna~Ortiz\,\href{https://orcid.org/0000-0002-7026-8198}{\includegraphics[height=1.7ex]{orcidlogo.png}}\textsuperscript{\,11}},
{D.\;R\"ohrich\textsuperscript{\,9}},
{E.\;Rondio\,\href{https://orcid.org/0000-0002-2607-4820}{\includegraphics[height=1.7ex]{orcidlogo.png}}\textsuperscript{\,13}},
{M.\;Roth\,\href{https://orcid.org/0000-0003-1281-4477}{\includegraphics[height=1.7ex]{orcidlogo.png}}\textsuperscript{\,4}},
{{\L}.\;Rozp{\l}ochowski\,\href{https://orcid.org/0000-0003-3680-6738}{\includegraphics[height=1.7ex]{orcidlogo.png}}\textsuperscript{\,12}},
{B.T.\;Rumberger\,\href{https://orcid.org/0000-0002-4867-945X}{\includegraphics[height=1.7ex]{orcidlogo.png}}\textsuperscript{\,24}},
{M.\;Rumyantsev\,\href{https://orcid.org/0000-0001-8233-2030}{\includegraphics[height=1.7ex]{orcidlogo.png}}\textsuperscript{\,20}},
{A.\;Rustamov\,\href{https://orcid.org/0000-0001-8678-6400}{\includegraphics[height=1.7ex]{orcidlogo.png}}\textsuperscript{\,1}},
{M.\;Rybczynski\,\href{https://orcid.org/0000-0002-3638-3766}{\includegraphics[height=1.7ex]{orcidlogo.png}}\textsuperscript{\,11}},
{A.\;Rybicki\,\href{https://orcid.org/0000-0003-3076-0505}{\includegraphics[height=1.7ex]{orcidlogo.png}}\textsuperscript{\,12}},
{D.\;Rybka\,\href{https://orcid.org/0000-0002-9924-6398}{\includegraphics[height=1.7ex]{orcidlogo.png}}\textsuperscript{\,13}},
{K.\;Sakashita\,\href{https://orcid.org/0000-0003-2602-7837}{\includegraphics[height=1.7ex]{orcidlogo.png}}\textsuperscript{\,7}},
{K.\;Schmidt\,\href{https://orcid.org/0000-0002-0903-5790}{\includegraphics[height=1.7ex]{orcidlogo.png}}\textsuperscript{\,16}},
{A.\;Seryakov\,\href{https://orcid.org/0000-0002-5759-5485}{\includegraphics[height=1.7ex]{orcidlogo.png}}\textsuperscript{\,20}},
{P.\;Seyboth\,\href{https://orcid.org/0000-0002-4821-6105}{\includegraphics[height=1.7ex]{orcidlogo.png}}\textsuperscript{\,11}},
{U.A.\;Shah\,\href{https://orcid.org/0000-0002-9315-1304}{\includegraphics[height=1.7ex]{orcidlogo.png}}\textsuperscript{\,11}},
{Y.\;Shiraishi\,\href{https://orcid.org/0000-0002-0132-3923}{\includegraphics[height=1.7ex]{orcidlogo.png}}\textsuperscript{\,8}},
{A.\;Shukla\,\href{https://orcid.org/0000-0003-3839-7229}{\includegraphics[height=1.7ex]{orcidlogo.png}}\textsuperscript{\,25}},
{M.\;S{\l}odkowski\,\href{https://orcid.org/0000-0003-0463-2753}{\includegraphics[height=1.7ex]{orcidlogo.png}}\textsuperscript{\,19}},
{P.\;Staszel\,\href{https://orcid.org/0000-0003-4002-1626}{\includegraphics[height=1.7ex]{orcidlogo.png}}\textsuperscript{\,14}},
{G.\;Stefanek\,\href{https://orcid.org/0000-0001-6656-9177}{\includegraphics[height=1.7ex]{orcidlogo.png}}\textsuperscript{\,11}},
{J.\;Stepaniak\,\href{https://orcid.org/0000-0003-2064-9870}{\includegraphics[height=1.7ex]{orcidlogo.png}}\textsuperscript{\,13}},
{F.\;Sutter\textsuperscript{\,4}},
{{\L}.\;\'Swiderski\,\href{https://orcid.org/0000-0001-5857-2085}{\includegraphics[height=1.7ex]{orcidlogo.png}}\textsuperscript{\,13}},
{J.\;Szewi\'nski\,\href{https://orcid.org/0000-0003-2981-9303}{\includegraphics[height=1.7ex]{orcidlogo.png}}\textsuperscript{\,13}},
{R.\;Szukiewicz\,\href{https://orcid.org/0000-0002-1291-4040}{\includegraphics[height=1.7ex]{orcidlogo.png}}\textsuperscript{\,18}},
{A.\;Taranenko\,\href{https://orcid.org/0000-0003-1737-4474}{\includegraphics[height=1.7ex]{orcidlogo.png}}\textsuperscript{\,20}},
{A.\;Tefelska\,\href{https://orcid.org/0000-0002-6069-4273}{\includegraphics[height=1.7ex]{orcidlogo.png}}\textsuperscript{\,19}},
{D.\;Tefelski\,\href{https://orcid.org/0000-0003-0802-2290}{\includegraphics[height=1.7ex]{orcidlogo.png}}\textsuperscript{\,19}},
{V.\;Tereshchenko\textsuperscript{\,20}},
{R.\;Tsenov\,\href{https://orcid.org/0000-0002-1330-8640}{\includegraphics[height=1.7ex]{orcidlogo.png}}\textsuperscript{\,2}},
{L.\;Turko\,\href{https://orcid.org/0000-0002-5474-8650}{\includegraphics[height=1.7ex]{orcidlogo.png}}\textsuperscript{\,18}},
{T.S.\;Tveter\,\href{https://orcid.org/0009-0003-7140-8644}{\includegraphics[height=1.7ex]{orcidlogo.png}}\textsuperscript{\,10}},
{M.\;Unger\,\href{https://orcid.org/0000-0002-7651-0272~}{\includegraphics[height=1.7ex]{orcidlogo.png}}\textsuperscript{\,4}},
{M.\;Urbaniak\,\href{https://orcid.org/0000-0002-9768-030X}{\includegraphics[height=1.7ex]{orcidlogo.png}}\textsuperscript{\,16}},
{D.\;Veberi\v{c}\,\href{https://orcid.org/0000-0003-2683-1526}{\includegraphics[height=1.7ex]{orcidlogo.png}}\textsuperscript{\,4}},
{O.\;Vitiuk\,\href{https://orcid.org/0000-0002-9744-3937}{\includegraphics[height=1.7ex]{orcidlogo.png}}\textsuperscript{\,18}},
{V.\;Volkov\,\href{https://orcid.org/0000-0002-4785-7517}{\includegraphics[height=1.7ex]{orcidlogo.png}}\textsuperscript{\,20}},
{A.\;Wickremasinghe\,\href{https://orcid.org/0000-0002-5325-0455}{\includegraphics[height=1.7ex]{orcidlogo.png}}\textsuperscript{\,22}},
{K.\;Witek\,\href{https://orcid.org/0009-0004-6699-1895}{\includegraphics[height=1.7ex]{orcidlogo.png}}\textsuperscript{\,15}},
{K.\;W\'ojcik\,\href{https://orcid.org/0000-0002-8315-9281}{\includegraphics[height=1.7ex]{orcidlogo.png}}\textsuperscript{\,16}},
{O.\;Wyszy\'nski\,\href{https://orcid.org/0000-0002-6652-0450}{\includegraphics[height=1.7ex]{orcidlogo.png}}\textsuperscript{\,11}},
{A.\;Zaitsev\,\href{https://orcid.org/0000-0003-4711-9925}{\includegraphics[height=1.7ex]{orcidlogo.png}}\textsuperscript{\,20}},
{E.\;Zherebtsova\,\href{https://orcid.org/0000-0002-1364-0969}{\includegraphics[height=1.7ex]{orcidlogo.png}}\textsuperscript{\,18}},
{E.D.\;Zimmerman\,\href{https://orcid.org/0000-0002-6394-6659}{\includegraphics[height=1.7ex]{orcidlogo.png}}\textsuperscript{\,24}}, and
{A.\;Zviagina\,\href{https://orcid.org/0009-0007-5211-6493}{\includegraphics[height=1.7ex]{orcidlogo.png}}\textsuperscript{\,20}}

\end{sloppypar}

\noindent
\textsuperscript{1}~National Nuclear Research Center, Baku, Azerbaijan\\
\textsuperscript{2}~Faculty of Physics, University of Sofia, Sofia, Bulgaria\\
\textsuperscript{3}~LPNHE, Sorbonne University, CNRS/IN2P3, Paris, France\\
\textsuperscript{4}~Karlsruhe Institute of Technology, Karlsruhe, Germany\\
\textsuperscript{5}~HUN-REN Wigner Research Centre for Physics, Budapest, Hungary\\
\textsuperscript{6}~E\"otv\"os Lor\'and University, Budapest, Hungary\\
\textsuperscript{7}~Institute for Particle and Nuclear Studies, Tsukuba, Japan\\
\textsuperscript{8}~Okayama University, Japan\\
\textsuperscript{9}~University of Bergen, Bergen, Norway\\
\textsuperscript{10}~University of Oslo, Oslo, Norway\\
\textsuperscript{11}~Jan Kochanowski University, Kielce, Poland\\
\textsuperscript{12}~Institute of Nuclear Physics, Polish Academy of Sciences, Cracow, Poland\\
\textsuperscript{13}~National Centre for Nuclear Research, Warsaw, Poland\\
\textsuperscript{14}~Jagiellonian University, Cracow, Poland\\
\textsuperscript{15}~AGH - University of Krakow, Poland\\
\textsuperscript{16}~University of Silesia, Katowice, Poland\\
\textsuperscript{17}~University of Warsaw, Warsaw, Poland\\
\textsuperscript{18}~University of Wroc{\l}aw,  Wroc{\l}aw, Poland\\
\textsuperscript{19}~Warsaw University of Technology, Warsaw, Poland\\
\textsuperscript{20}~Affiliated with an institution covered by a cooperation agreement with CERN\\
\textsuperscript{21}~University of Belgrade, Belgrade, Serbia\\
\textsuperscript{22}~Fermilab, Batavia, USA\\
\textsuperscript{23}~University of Notre Dame, Notre Dame, USA\\
\textsuperscript{24}~University of Colorado, Boulder, USA\\
\textsuperscript{25}~University of Hawaii at Manoa, Honolulu, USA\\
\textsuperscript{26}~University of Pittsburgh, Pittsburgh, USA\\
\textsuperscript{27}~University of Geneva, Geneva, Switzerland\footnote{No longer affiliated with the NA61/SHINE collaboration}\\

\appendix

\section{Beam Selection and Formalism of the Analysis}
\label{appendix}

\subsection{Auxillary Measurements of Beam Particles}
\label{app:A1}

The online carbon trigger ($\Zsq=36$) allows the neighboring elements and their corresponding isotopes up to a charge difference, $\Delta Z = 1$ and mass difference $\Delta A = 2$ nuclei.
This provides an excellent opportunity to make additional measurements by selecting these isotopes as the primary beam particle.
These auxiliary measurements are done by altering the upstream cuts on the calibrated data to select a desired nucleus as the primary beam particle, such as the isotopes neighboring the \iso{12}{C}.
The measurements are necessary for calculating various corrections and systematics as described in \cref{app:B}.
For instance, in order to study feed-down reactions of the type, A+p$\rightarrow$f, where A$\neq$\iso{12}{C} and corresponds to a neighboring nucleus producing a fragment nucleus, \f.
These reactions contribute to the true signal from \iso{12}{C}+p reactions which needs to be corrected for.
Equivalently, it also introduces a systematic uncertainty on the measurement (see \cref{sec:imp} and \cref{app:A2} for further details).
The auxiliary measurements are also used to calculate the mass-changing cross section of a desired nucleus, which can then be input to correct for in-target interactions of this nucleus with the target nucleus, leading to a loss of the expected signal (detailed in \cref{sec:recIntTar}).

For our measurement, we are interested in the feed-down cross section of the following nuclei: \iso{13}{C}, \iso{11}{C}, \iso{15}{N}, \iso{14}{N}, \iso{11}{B}, and \iso{10}{B}.
These fragments are present in the data (\cref{fig:beamComp}) and are selected by placing appropriate cuts on the time-of-flight as measured between the A and the S1 scintillators, and the charge as measured from the energy loss in the S1 detector, $Z^2_\text{S1}$.
Furthermore, all other upstream cuts listed in \cref{tab:1} applied for \iso{12}{C} selection remain unchanged.
Instead of the 2D Gaussian with tails function used to select the \iso{12}{C} beam particle (see~\cref{sec:data}), simple rectangular cuts were applied to select a particular nucleus by placing an upper and lower limit on the time-of-flight and $Z^2_\text{S1}$.
The total number of beam particles corresponding to these nuclei for the three target settings, PE, C, and OUT are given in \cref{tab:A1}.

\begin{table}
	\caption{The statistics for each of the nuclei selected as the
        primary beam particle. The column labeled $N_\text{b}$ gives
        the total number of events, (number of this nucleus in the
        beam) recorded for every nucleus $i$ and the numbers of these
        nuclei as measured in the MTPC are listed in the $N_\text{i}$
        column. The total measured mass-changing probability is given in he $P^\mathrm{IN/OUT}_\mathrm{i\to X}$, and the last column gives the target-out-subtracted mass-changing probability, inside the target T.}  \label{tab:A1}
\begin{center}
	\begin{tabular}{l l r r r r r}
	\toprule
	Nucleus (i) & Target & $N_\text{b}$ & $N_\text{i}$ & $\PP{IN/OUT}{i}{X} = \left(1 - \frac{N_\text{i}}{N_\text{b}} \right)$ & $P^\text{T} = \left(\frac{P^\text{IN} - P^\text{OUT}}{1 - P^\text{OUT}}\right)$\\
	\midrule
				& PE & 35810 & 30347 & $0.152\pm0.002$ & $0.084\pm0.003$ \\
	\iso{13}{C} & C & 31501 & 26887 & $0.146\pm0.002$ & $0.077\pm0.004$ \\
				& OUT & 7953 & 7357 & $0.075\pm0.003$ & - \\
	\midrule
				& PE & 12300 & 10595 & $0.139\pm0.003$ & $0.068\pm0.006$ \\
	\iso{11}{C} & C & 10490 & 9144 & $0.128\pm0.003$ & $0.057\pm0.006$ \\
				& OUT & 2669 & 2466 & $0.076\pm0.005$ & - \\
	\midrule
				& PE & 1692 & 1191 & $0.296\pm0.011$ & $0.098\pm0.030$ \\
	\iso{15}{N} & C & 1292 & 949 & $0.265\pm0.012$ & $0.059\pm0.030$ \\
				& OUT & 315 & 246 & $0.219\pm0.023$ & - \\
	\midrule
				& PE & 1607 & 1302 & $0.190\pm0.098$ & $0.102\pm0.019$ \\
	\iso{14}{N} & C & 1383 & 1151 & $0.168\pm0.010$ & $0.078\pm0.020$ \\
				& OUT & 349 & 315 & $0.097\pm0.016$ & - \\
	\midrule
				& PE & 1929 & 1379 & $0.285\pm0.010$ & $0.061\pm0.028$ \\
	\iso{11}{B} & C & 1743 & 1261 & $0.276\pm0.011$ & $0.049\pm0.028$\\
				& OUT & 473 & 360 & $0.239\pm0.020$ & - \\
	\midrule
				& PE & 3561 & 2872 & $0.193\pm0.007$ & $0.109\pm0.012$ \\
	\iso{10}{B} & C & 3100 & 2581 & $0.167\pm0.007$ & $0.081\pm0.013$ \\
				& OUT & 785 & 711 & $0.094\pm0.010$ & - \\
	\bottomrule
	\end{tabular}
 \end{center}
\end{table}

\subsection{Feed-down due to Beam Impurities}
\label{app:A2}

\def\Ni{$N_$}
\def\Nb{$N_\mathrm{b}$\xspace}
\def\Nfm{$N_\mathrm{f}^\mathrm{m}$}
\def\i{`i'\xspace}

The upstream cut defined in \cref{sec:data} is optimized to select \iso{12}{C} as the primary beam particle.
Nevertheless, the shape of the fragment distribution in the time-of-flight vs $Z^2_\text{S1}$ plot allows the neighboring isotopes to be selected into the cut.
These beam nuclei interact with the target material producing boron and carbon fragments studied in our analysis.
Let \Nb be the total number of particles constituting the beam, and let $N_{^{12}\text{C}}$ be the number of \iso{12}{C} nuclei present after the upstream selection.
The number of a particular type of beam nuclei is estimated from the fit performed to the upstream distribution of fragments (\cref{app:A4}).
Then the fraction of \iso{12}{C} nucleus is given by, $f_{^{12}\text{C}}=\frac{N_{^{12}\text{C}}}{N_\text{b}}$, and the fraction of the nucleus i, neighboring \iso{12}{C} is given by, $f_\text{i} = \frac{N_\text{i}}{N_\text{b}}$.
Then it immediately follows that:
\begin{subequations}
\begin{equation}
f_\text{tot}=\sum_\text{i} f_\text{i} + f_{^{12}\text{C}}=1
\label{eq:A11}
\end{equation}
\begin{equation}
N_\text{b} = N_{^{12}\text{C}} + \sum_\text{i} N_\text{i}
\label{eq:A12}
\end{equation}
\end{subequations}
Let $N_\text{f}^\text{m}$ be the total number of fragments of type \f produced in the beam-target interaction and as measured in the MTPC.
Then this number is written as a sum of the number fragments produced by all the beam nuclei as:
\begin{equation}
    N_\text{f}^\text{m} = N_{^{12}\text{C}\to{f}} + \sum_\text{i} N_{\text{i}\to\text{f}}.
    \label{eq:A2}
\end{equation}
The quantity $N_{\text{i}\to\text{f}}$ is the number of nucleus \f produced as a result of the fragmentation of nucleus \i.
This number can, therefore, be expressed as the product of the corresponding production probability and the total number of nucleus i available for the interaction, as, $N_{\text{i}\to\text{f}}=\Ps{i}{f}\,N_\text{i}$.
This also holds for \f produced by fragmenting \iso{12}{C}.
Thus we can divide \cref{eq:A2} on both sides by the total number of beam particles, \Nb to obtain the measured production probability $P_\text{f}^\text{m}$, as,
\begin{equation}
P_\text{f}^\text{m} =
  \frac{N_\text{f}^\text{m}}{N_\text{b}} =
    \frac{N_{^{12}\text{C}}\,P_{^{12}\text{C}\to\text{f}} + \sum_\text{i} N_\text{i}\,\Ps{i}{f}}
         {N_\text{b}}.
\label{eq:A3}
\end{equation}
The probability $P_{^{12}\text{C}\to\text{f}}$ on the right side of \cref{eq:A3} is the true production probability of nucleus \f from \iso{12}{C}+T interactions.
Therefore, making the necessary substitutions for the fraction of beam nuclei, and solving \cref{eq:A3} for the true probability, we obtain the final expression:
\begin{equation}
P_{^{12}\text{C}\to\text{f}} =
  \frac{P_\text{f}^\text{m} - \sum_\text{i} f_\text{i}\,\Ps{i}{f}}{f_{^{12}\text{C}}}.
\label{eq:A4}
\end{equation}
The feed-down probabilities $\Ps{i}{f}$ of the impurity nuclei \i in \cref{eq:A4} are measured quantities that can be calculated explicitly by altering the upstream cuts.
This enables us to select the nucleus \i as the primary beam particle and compute the production of the nucleus \f.
The relative fraction $f_\text{i}$ of the nuclei neighboring \iso{12}{C} are given in \cref{tab:A2}.

\begin{table}
\caption{Relative fractions of the number of impurity nuclei events present in the upstream selection of neighboring \iso{12}{C} nucleus.}
\label{tab:A2}
\begin{center}
\begin{tabular}{l r}
	\toprule[0.06em]
	Impurity (i) & fraction $f_\text{i}$ \\
	\midrule[0.06em]
    \iso{15}{N} & $6.7{\times}10^{-4}$\\
	\iso{14}{N} & $1.7{\times}10^{-3}$\\
	\iso{13}{C} & $8.0{\times}10^{-6}$\\
	\iso{11}{C} & $2.1{\times}10^{-3}$\\
	\iso{11}{B} & $7.0{\times}10^{-5}$\\
    \midrule[0.06em]
     & $\sum_\text{i} f_\text{i}=4.5{\times}10^{-3}$\\
    \bottomrule[0.06em]
\end{tabular}
\end{center}
\end{table}

The systematic uncertainty resulting from this correction is estimated from the statistical uncertainty of these auxiliary measurements.
\cref{eq:A4} is used to calculate the corrected production probability of the fragments \iso{11}{C}, \iso{11}{B}, and \iso{10}{B} given in \cref{sec:imp}.
The correction on the cross section and the resulting systematic is sub-dominant and at the order of ${\sim}1\%$ of the measured cross section of the isotopes.

\subsection{Inelastic Re-interaction inside the Target}
\label{app:A3}

Consider a beam composed of a single type of nucleus, \b, and let $N_\text{b}$ be the total number of beam particles incident on the target (T) of thickness $d$.
Then the rate of inelastic interactions per unit length interval $\dd x$ of the target is given by:
\begin{equation}
	\frac{\dd N_\text{b}(x)}{\dd x} = -\frac{N_\text{b}(x)}{\lambda_\text{b}},
 \label{A2:eq1}
\end{equation}
where $\lambda_\text{b}$ is the interaction length of nucleus \b for a given target material and the negative sign denotes the destruction of particles.
If the nucleus \b fragments to \f, then the differential equation describing the production and destruction of nucleus type \f in a length interval $\dd x$ is given by:
\begin{equation}
\frac{\dd N_\text{f}(x)}{\dd x} = \frac{N_\text{b}(x)}{\lambda_{\text{b}\to\text{f}}} - \frac{N_\text{f}(x)}{\lambda_\text{f}}.
\label{A2:eq2}
\end{equation}
The second term denotes the inelastic interaction of nucleus \f inside the target, and, hence, additional measurements of nucleus \f as primary beam particles are required for this calculation.
This is an implicit assumption and is facilitated by the auxiliary measurements described in \cref{app:A1}.
In a more general form, \cref{A2:eq2} can be expanded to a set of linear differential equations which can be written using vector notation as:
\begin{equation}
\frac{\dd \vec{N}(x)}{\dd x} = M \, \vec{N}(x).
\label{A2:eq3}
\end{equation}
Here, $\vec{N}(x)$ is a vector of the number of particles at any given distance $x$ inside the target.
For convenience of notation, we introduce a matrix $M$ comprising of the inverse of the interaction lengths relevant for the production of a given nuclei \f.
Then the solution to the first order linear differential \cref{A2:eq1,A2:eq3} gives the number of particles $N_\text{b}$ and $N_\text{f}$ respectively, at a given distance $x$ inside the target, and is written as,
\begin{equation}
\vec{N}(x) = \vec{N}(0)\exp(Mx).
\label{A2:eq4}
\end{equation}
Here $\vec{N}(0)$ is the number of particles incident on the surface
of the target, at $x=0$.  For an inelastic interaction of a nucleus \b
inside the target of thickness d, the matrix $M$ contains a single
element $M_\text{bb}=-1/\lambda_\text{b}$ corresponding to the
interaction length of nucleus \b.  Hence,~\cref{A2:eq4} reduces to the
following form:
\begin{equation}
    N_\text{b}(x) = N_\text{b}(0)\exp{(-x/\lambda_\text{b})}.
    \label{A2:eq5}
\end{equation}
This equation can be solved for the matrix element $M_\text{bb}$ by re-arranging the terms and taking natural logarithms on both sides, resulting in the final expression:
\begin{equation}
	-\frac{1}{\lambda_\text{b}} = M_\text{bb} = \frac{1}{x}\ln\frac{N_\text{b}(x)}{N_\text{b}(0)}.
	\label{A2:eq6}
\end{equation}
Let us consider the reaction b$\rightarrow$f concerning the production of \f from nucleus \b interacting inside the target material.
In this case, the matrix $M$ is a lower triangular matrix containing the inelastic interaction lengths of nuclei \b and \f as well as the interaction length leading to the production of \f from \b, $\lambda_{\text{b}\to\text{f}}$.
Therefore, \cref{A2:eq4} can be written as:
\begin{equation}
\begin{pmatrix}
    N_\text{b}(x) \\
    N_\text{f}(x) \\
\end{pmatrix} = \exp\begin{pmatrix}
-1/\lambda_\text{b} & 0 \\
1/\lambda_{\text{b}\to\text{f}} & -1/\lambda_\text{f} \\
\end{pmatrix}
\begin{pmatrix}
    N_\text{b}(0)\\
    0 \\
\end{pmatrix}
.
\label{A2:eq7}
\end{equation}
The left-hand side in the equation above represents the vector of the number of nuclear particles \b and \f at a distance $x$ inside the target, whereas the beam composition is represented by the column vector on the right.
We have assumed a pure beam consisting of only a single type of nucleus \b, the initial number of which is $N_\text{b}(0)$, while in principle, the presence of beam impurities is inevitable.
The quantity of interest for our measurement is $\lambda_{\text{b}\to\text{f}}$ which can be determined by evaluating the exact solution of the exponential form of the matrix $M$ and is written as,
\begin{equation}
\frac{1}{\lambda_\text{bf}} =
  \frac{M_\text{ff} - M_\text{bb}}{\exp(M_\text{ff}x) - \exp(M_\text{bb}x)}
  \,
  \frac{N_\text{f}(x)}{N_\text{b}(0)}.
\label{A2:eq8}
\end{equation}
The expression for $\lambda_\text{bf}$ derived above requires additional measurement of the inelastic interaction probability of nucleus \f denoted by the matrix element $M_\text{ff}$.
It is computed similarly as for nucleus \b using \cref{A2:eq6}. The expression for the true in-target interaction length for nucleus \f given in \cref{A2:eq8} is then used to correct the calculated production cross section of the isotopes, where $\text{f}=\text{\iso{11}{C}}$, \iso{11}{B}, and \iso{10}{B}, as given in \ref{sec:recIntTar}.

\subsection[Offline Selection of 12C as the Beam Particle]{Offline Selection of \iso{12}{C} as the Beam Particle}
\label{app:A4}

In order to select \iso{12}{C} as the primary beam particle, we make an offline selection on the triggered data (\cref{fig:beamComp}) by fitting the \tof vs $Z^2_\text{S1}$ distribution of the beam composition.
An optimal cut is then made based on the fit, by maximizing the total number of \iso{12}{C} events and simultaneously keeping the neighboring isotopes at a minimum.
A 2-dimensional Gaussian with exponential tails function is used for the fit, to describe the drop-shaped isotope peaks as seen in the \tof vs.\ $Z^2$ plot of \cref{fig:beamComp}.
This function can be mathematically described as a sum of two 1-dimensional Gaussian with exponential tails, $f(x)$ and $g(y)$ along the $x$ and $y$ axes corresponding to the $Z^2$ and \tof axes respectively as:
\begin{equation}
	F(x,y) = N\left( f(x) + g(y) \right),
	\label{eq:A41}
\end{equation}
where $N$ is the normalization of each of the peaks, and the functions $f(x)$ and $g(y)$ are written as:
\begin{subequations}
\begin{equation}
	f(x) = \frac{1}{2} \left[ \exp\left(-\lambda_x\left(x-Z^2-\frac{\sigma_x^2\lambda_x}{2}\right)\right)\left(1 + \erf(a(x))\right)\right],
\end{equation}
\begin{equation}
	g(y) = \frac{1}{2} \left[ \exp\left(-\lambda_y\left(y-t^2-\frac{\sigma_y^2\lambda_y}{2}\right)\right)\left(1 + \erf(b(y))\right)\right],
\end{equation}
\label{eq:A42}
\end{subequations}
where $Z^2$ is the squared charge as measured in S1 and $t$ is the time of flight difference between the A and the S1 scintillators.
The parameter $\sigma$ corresponds to the width of the Gaussian function whereas $\lambda$ parameterizes the exponential tails.
The arguments of the Error functions, $a(x)$ and $b(y)$, in the equations above are also functions of $\sigma$ and $\lambda$, and are expressed as:
\begin{subequations}
	\begin{equation}
		a(x) = \left( x - Z^2 - \frac{\sigma_x^2}{\lambda_x}\right)\frac{1}{\sigma_x\sqrt{2}},
	\end{equation}
	\begin{equation}
		b(y) = \left( y - t^2 - \frac{\sigma_y^2}{\lambda_y}\right)\frac{1}{\sigma_y\sqrt{2}}.
	\end{equation}
\label{eq:A43}
\end{subequations}

\begin{table}[t]
\caption{Function parameter values used to select primary \iso{12}{C} nucleus to make the optimal cut at $k\approx10\%$, as described in the text.}
\label{tab:A4}
\begin{center}
\begin{tabular}{c c c c c c}
\toprule
$Z^2$ & $\sigma_x$ & $\lambda_x$ & $t$ & $\sigma_y$ & $\lambda_y$ \\
\midrule
35.00 & 1.78 & 0.71 & -0.04 & 0.04 & 19.83 \\
\bottomrule
\end{tabular}
\end{center}
\end{table}

\begin{figure}
	\centering
	\includegraphics[width=0.84\linewidth]{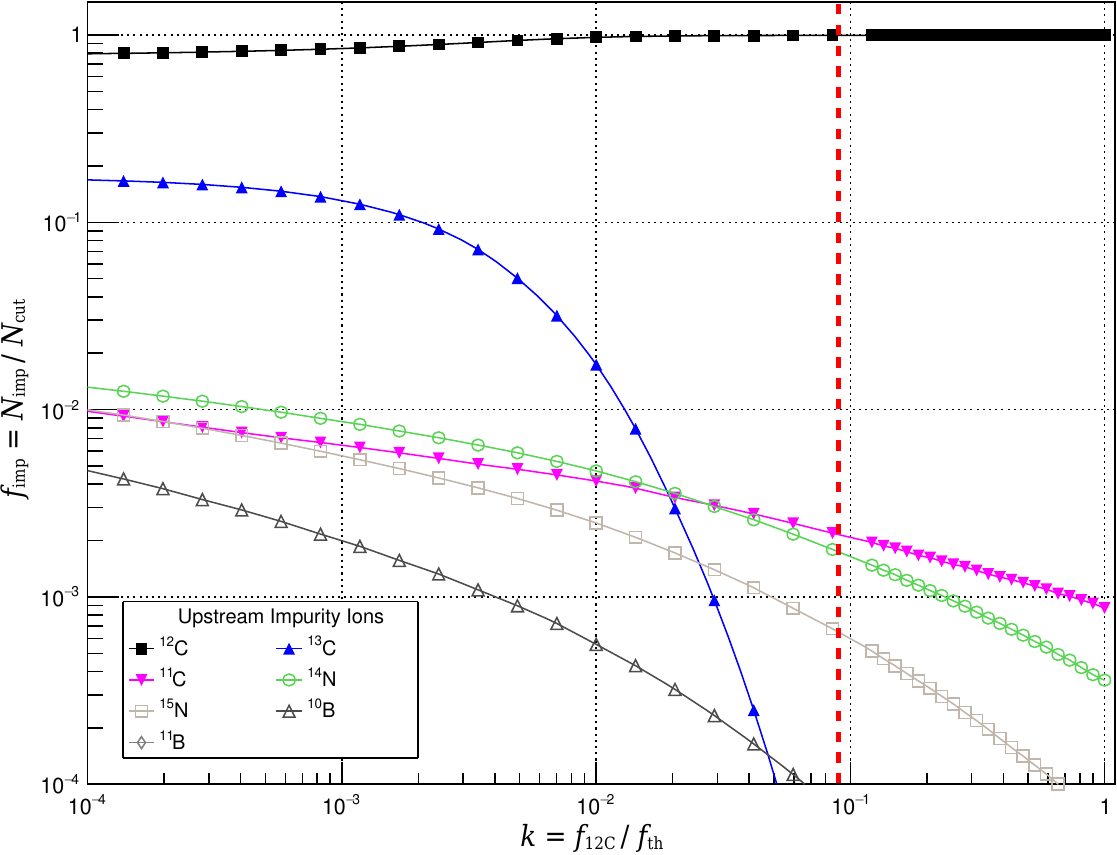}
	\caption{The lines show the fraction of the neighboring isotopes, $f_\text{imp}$ in the \iso{12}{C} cut as a function of the fraction of \iso{12}{C} in the cut ($k$), shown on the $x$-axis.
 The vertical dashed red line at $k{\sim}10^{-1}$ is the optimal cut level for \iso{12}{C}, used in this analysis (see text for details).}
	\label{fig:11}
\end{figure}

A numerical procedure is undertaken to determine the optimal cut for the analysis.
We begin by defining the term, relative fraction of the \iso{12}{C} peak $k$, as the ratio of the maximum of the fit \iso{12}{C} peak $f_{^{12}\text{C}}$ to the set threshold value $f_\text{th}$, as,
\begin{equation*}
	k = \frac{f_\text{12C}}{f_\text{th}},
\end{equation*}
where $f_\text{th}$ is a free parameter used for optimization, and the ratio $k=1$ simply corresponds to the pure \iso{12}{C} peak, with a very small number of events and negligible contribution from the neighboring isotopes in the cut (see \cref{fig:11}).
Next, we determine the number of nuclear fragments produced in beam-target interactions ($N$), for instance, the production of \iso{11}{C}, and compute its relative uncertainty.
This procedure is repeated for decreasing values of $k$, which means moving farther away from the \iso{12}{C} peak and widening the cut to gather more statistics.
At $k\approx1$, the relative uncertainty is dominated by the statistical error ($\upsigma{N}_\text{stat.}$) owing to a few events, whereas as $k$ decreases, the systematic uncertainty ($\upsigma{N}_\text{syst.}$) increases due to the presence of the neighboring isotopes.
The optimum is defined as the value of $k$ where the relative error on N is minimal.
This corresponds to $k\approx0.1$ and is the same for measuring the production of all the fragments undertaken in this work.
The cut corresponding to this optimal value of $k$ is shown as a black contour in \cref{fig:beamComp}.

\subsection{Calculation of the Charge-changing Probability}
\label{sec:ccXsec}

A charge-changing reaction is defined as an interaction where a nucleus of charge $Z_\text{B}$ fragments to any nucleus of charge $Z_\text{Y}$ after interacting with the target (T), such that $\Delta{Z} = Z_\text{B} - Z_\text{Y} \geqslant 1$.
To calculate the total charge-changing cross section of the nucleus `B' we select all tracks in the MTPC corresponding to charge $Z_\text{B}$, and denote the reaction as $\text{B}\to\text{Y}$.
Let $N_\text{beam}$ be the total number of beam particles incident on the target and $N_\text{B}$ be the total number of `B' tracks as measured in the MTPC.
Then, following the notation introduced in \cref{sec:analysis}, the total measured survival probability of a \b nucleus, is written as, $\Ps{B}{B} = N_\text{B}/N_\text{beam}$.
Therefore, the total measured charge-changing probability is then, $\Ps{B}{Y} = 1 - \Ps{B}{B}$.
It can be expressed as a product of the survival probabilities of a carbon nucleus in the four regions described in the previous section and equivalent to \cref{eq:1} for the IN and~\cref{eq:2} for the OUT case.
On solving these equations for the charge-changing case, we can write the true interaction probability in the target similar to \cref{eq:3}, as:
\begin{equation}
\PP{T}{B}{Y} =
  \left(1 - \PP{T}{B}{B}\right) = \left(1 - \frac{\PP{IN}{B}{B}}{\PP{OUT}{B}{B}}\right).
\end{equation}
The probabilities in the ratio term on the right are direct measurable quantities from the MTPC measurement for the IN and OUT cases.
To determine the charge-changing probability of a \iso{12}{C}, we simply replace $N_\text{B}$ with the measured number of tracks $N_\text{C}$ in the MTPC corresponding to $Z=6$ particles.

\subsection{Calculation of the Probability for Isotopic Production}
\label{sec:isoXsec}

To derive the probability for in-target production of a particular nucleus \f resulting from fragmentation of beam nucleus \b, we work with the total probability measured with the target (IN).
Let us consider the simplest case of \iso{12}{C} nucleus fragmenting to \iso{11}{C}, where the difference in the mass number of nucleus \f and \b is 1, that is, $\Delta A = A_\text{b}-A_\text{f} = 1$.
Then, the total measured production probability is expressed by the following equation:
\begin{equation}
\PP{IN}{b}{f} =
  \PP{up}{b}{f} \, \PP{T}{f}{f} \, \PP{VD}{f}{f} \, \PP{dn}{f}{f} +
  \PP{up}{b}{b} \, \PP{T}{b}{f} \, \PP{VD}{f}{f} \, \PP{dn}{f}{f} +
  \PP{up}{b}{b} \, \PP{T}{b}{b} \, \PP{VD}{b}{f} \, \PP{dn}{f}{f}.
\label{eq:4}
\end{equation}
The first term denotes the production of nucleus \f upstream of the target that survives an inelastic interaction with the target and any of the downstream detectors, whereas the second term denotes the production of \f inside the target provided the beam nucleus \b does not interact upstream, and the produced \f survives the downstream region before it reaches the MTPC.
The third term denotes the probability of the beam nucleus interacting very close to but downstream of the target.
The nucleus \f produced by the fragmentation of \b upstream of the target, at the target, and also downstream, very close to the target share the same relative position $\Delta x$ at the end of the MTPC.
Hence the total number of \f fragments measured in the MTPC when the target is removed (OUT) can be expressed as the product of upstream and close-to-target interactions of the beam, provided that the nuclear fragment \f does not interact inside the magnetic field, resulting in loss of the signal.
The total measured probability for the empty target holder (OUT) is written as:
\begin{equation}
\PP{OUT}{b}{f} =
  \PP{up}{b}{f} \, \PP{VD}{f}{f} \, \PP{dn}{f}{f} +
  \PP{up}{b}{b} \, \PP{VD}{b}{f} \, \PP{dn}{f}{f}.
\label{eq:5}
\end{equation}
The fragments \f produced downstream of the target, inside the magnetic field are deflected to a lesser extent relative to the beam, as opposed to the true signal of the fragments produced at the target.
These nuclei do not traverse completely through the magnetic field and possess a smaller spread in the longitudinal momentum resulting from the Fermi motion of the fragment, and thus a smaller width $\sigma_\text{F}$.
This enables us to distinguish these downstream fragments from the true \f signal in the fit (blue peak in \cref{fitC}).
Therefore, it does not contribute to the total \f production probability inside the target, $\left(\PP{IN}{b}{f}\right)$.
To determine the in-target production probability we begin with subtracting \cref{eq:5} from \cref{eq:4}.
The resultant expression is:
\begin{equation}
\PP{IN}{b}{f} - \PP{OUT}{b}{f} =
  \PP{up}{b}{f} \, \PP{VD}{f}{f} \, \PP{dn}{f}{f} \left(\PP{T}{f}{f} - 1\right) +
  \PP{up}{b}{b} \, \PP{VD}{b}{f} \, \PP{dn}{f}{f} \left(\PP{T}{b}{b} - 1\right) +
  \PP{up}{b}{b} \, \PP{T}{b}{f} \, \PP{VD}{f}{f} \, \PP{dn}{f}{f}.
\label{eq:6}
\end{equation}
To solve this equation for a desired target T, we need to determine the unknown probabilities from auxiliary measurements of nucleus \f (see \cref{app:A1} for details).
We begin by writing \cref{eq:2} for \f and substituting the product term $\PP{VD}{f}{f}\,\PP{dn}{f}{f}$ in \cref{eq:6} with the ratio:
\begin{equation}
\PP{VD}{f}{f} \, \PP{dn}{f}{f} =
  \frac{\PP{OUT}{f}{f}}{\PP{up}{f}{f}}.
\label{eq:7}
\end{equation}
Next, we re-arrange the terms in \cref{eq:5} to obtain:
\begin{equation}
\PP{up}{b}{b} \, \PP{VD}{b}{f} \, \PP{dn}{f}{f} =
  \left(\PP{OUT}{b}{f} - \PP{up}{b}{f} \, \PP{VD}{f}{f} \, \PP{dn}{f}{f}\right).
\label{eq:8}
\end{equation}
The left-hand side expression in the above equation is then used to replace the product of probabilities in the second term in \cref{eq:6}.
After making the necessary substitutions from \cref{eq:7} and \cref{eq:8} and re-arranging the terms to solve for the in-target production probability of \f, we obtain its final expression,
\begin{equation}
\PP{T}{b}{f} =
  \frac{\PP{up}{f}{f}}
       {\PP{up}{b}{b}}
  \left(
    \frac{1}{\PP{OUT}{f}{f}}
      \left(
        \PP{IN}{b}{f} - \PP{OUT}{b}{f}\,\PP{T}{b}{b}
      \right)
  \right) -
  \frac{\PP{up}{b}{f}}
       {\PP{up}{b}{b}}
    \left(\PP{T}{f}{f} - \PP{T}{b}{b}\right).
\label{eq:9}
\end{equation}
All the probabilities on the right side of \cref{eq:9} are direct measurable quantities except for the leading ratio terms concerning the upstream probabilities.
These probabilities can be estimated by identifying the material of the beam-line detectors.
The S1 scintillator situated approximately 36\,m upstream of the target is a 0.5\,cm thick sheet of a polymer called vinyl toluene (BC-408).
Its physical properties make it remarkably similar to the polyethylene target (PE) and render it an appropriate site for the interaction of the beam particles.
Therefore, the upstream survival probability of any nucleus (\b or \f) can be estimated from its mass-changing probability with the PE target, as given in \cref{eq:3}.
Further details on the derivation of this quantity are explained in the \cref{app:A5}.
The second term in \cref{eq:9} contains the upstream production probability of \f from the fragmentation of \b.
Using the same analogy as before, the total measured production probability, $\PP{IN}{b}{f}$ can be used to set an upper bound on this quantity.
Whereas the denominator in the second term is the upstream survival probability of \b.
The quantity $\PP{IN}{b}{f}$ is relatively small and amounts to only ${\approx}0.5$\% of the measured $\PP{T}{b}{b}$.

It is crucial to note that the second term in the above~\cref{eq:9} is ${\leqslant}1$\% of the first term.
The final expression given in \cref{eq:9} applies for the case, $\Delta A=1$, e.g.,\ the production of \iso{11}{C} and \iso{11}{B} from the fragmentation of the \iso{12}{C} nucleus.
The production of $\Delta A = 2$ isotopes like \iso{10}{B} and \iso{10}{C} from \iso{12}{C}+T interaction proceeds analogous to the analysis described in this section.
With the exception that it includes additional contributions from indirect production channels as well.
Beam fragmentation upstream of the target can lead to production of $\Delta A = 1$ particles, which can then feed down to the production of the required $\Delta A = 2$ nuclei when interacting with the target, for instance, \iso{12}{C}+S1$\to$\iso{11}{C}+T$\to$\iso{10}{B}.
Let us once again denote the fragmenting nucleus of interest, by \b, and the fragment produced by \f, where \f = $\Delta A = 2$ nuclear fragments.
Then the total measured probability in this case, when the target is inserted (IN) is similar to \cref{eq:5}, and is written as
\begin{equation}
\PP{IN}{b}{f} =
  \PP{up}{b}{f} \, \PP{T}{f}{f} \, \PP{VD}{f}{f} \, \PP{dn}{f}{f} +
  \sum_\text{i}\left(\PP{up}{b}{i} \, \PP{T}{i}{f}\right)
  \PP{VD}{f}{f} \, \PP{dn}{f}{f} + \PP{up}{b}{b} \, \PP{T}{b}{f} \, \PP{VD}{f}{f} \, \PP{dn}{f}{f} +
  \PP{up}{b}{b} \, \PP{T}{b}{b} \, \PP{VD}{b}{f} \, \PP{dn}{f}{f}.
\label{eq:10}
\end{equation}
The first, third, and fourth terms in the equation above denote the upstream, in-target, and close-to but downstream of the target production of \f from \b respectively, and are exactly similar to \cref{eq:4}.
The second term indicates a sum over the in-target production probabilities of \f from the intermediate $\Delta A = 1$ nuclei (denoted as `i'), interacting with the target.
The second term is a two-step process leading to \f as b$\to$i$\to$f.
As the momentum per nucleon, $p_A$ is conserved in the process of fragmentation, fragments \f from the i+T interaction are deflected inside the vertex magnets in the $x$-$z$ plane exactly as the true signal from b+T interaction.
Nevertheless, the product of the probabilities in the second term for the two-step reaction (b$\to$i$\to$f) is $\mathcal{O}(P^2)\approx10^{-6}$ and accounts for just 0.1\% of $\PP{T}{b}{f}$ (see~\cref{app:A5} for further details).
This simplifies our analysis to the same formalism developed for $\Delta A= 1$ fragments in \cref{eq:4}.
Similarly, the total measured probability for the OUT case and eventually the final expression for in-target production of $\Delta A = 2$ fragments are written exactly as \cref{eq:5} and \cref{eq:9} respectively.
As in the case of \iso{11}{C} and \iso{11}{B}, the auxiliary measurements from primary \iso{10}{B} data present in the beam composition, are used in this analysis.

\subsection{Interaction of the Beam Upstream of the Target}
\label{app:A5}

The various detectors placed on the beamline have to be considered to evaluate if the beam particles interacted before the target.
Beam line detectors placed before the target constitute the upstream region of the experiment, from the scintillator S1 up to BPD-3, and are used for characterizing and studying the properties of the beam.
The counting detector S1 is an organic scintillator made from polyvinyl toluene, with a density, $\rho = 1.03$\,g/cm$^3$.
The scintillator has dimensions $L{\times}B{\times}T = 6{\times}6{\times}0.5$\,cm$^3$.
Its physical properties are similar to the PE target and has the highest upstream material budget compared to the BPDs and air.
The S1 is crucial for defining the trigger logic for online tagging of carbon particles and hence cannot be excluded from the beam line.
Therefore, the incoming beam nuclei can interact with S1, which is to say that the upstream survival probability of a beam nucleus \b, $\PP{up}{b}{b} \neq 1$.
However, since we can measure the mass-changing probability of any nucleus interacting with the PE and C targets (see \cref{app:A1}), we can utilize these values to compute the interaction probability with S1, or conversely the survival probability of any beam nuclei.

Following the analysis detailed in \cref{sec:analysis}, the mass-changing cross section ($\sigma^\text{T}_{\text{b}\to\text{X}}$) on a proton target for a nucleus \b can be computed.
Similar to the case of polyethylene, given that the carbon-to-hydrogen ratio for each polymer cell in vinyl toluene is known to be $\text{C:H}=1:1.1$, the mass-changing cross section of \b interacting with S1 can be computed with:
\begin{equation}
\sigma^\text{S1}_{\text{b}\to\text{X}} =
  \sigma^\text{C}_{\text{b}\to\text{X}} + 1.1\sigma^\text{p}_{\text{b}\to\text{X}}.
    \label{eq:A52}
\end{equation}
In general, the survival probability of the nucleus \b for a target T is written as, $\PP{T}{b}{b}= (1 - \PP{T}{b}{X})$ and the interaction probability is expressed in terms of the cross sections and target parameters as, $\PP{T}{b}{X} = 1 - \exp{\left(-d_\text{T}\,n_\text{T}\,\sigma^\text{T}_{\text{b}\to\text{X}}\right)}$, where $d_\text{T}$ is the target thickness and $n_\text{T}$ is the number density of the target.
For exponent values $x{\ll}1$, the expression $f(y) = 1 - \exp{(-y)} \approx y$.
With this approximation, we get $\PP{T}{b}{X} \approx d_\text{T}\,n_\text{T}\,\sigma^\text{T}_{\text{b}\to\text{X}}$ implying that $\PP{T}{b}{b} \approx (1 - d_\text{T}\,n_\text{T}\,\sigma^\text{T}_{\text{b}\to\text{X}})$.
This enables us to re-substitute for the ratio of the upstream survival probabilities of the \f and \b nuclei in \cref{eq:9}.

\begin{table}
    \caption{The mass-changing cross section of the primary \iso{11}{C}, \iso{11}{B}, and \iso{10}{B} nucleus on the C and p target.
    The cross section on S1 is calculated as per~\cref{eq:A52}.}
    \label{tab:a51}
    \begin{center}
    \resizebox{\linewidth}{!}{\begin{tabular}{
    	c c l l l l r
    }
    \toprule
        Label & Beam (i) & {$\sigma^\text{C}_{\text{i}\to\text{X}}$ (barn)} & {$\sigma^\text{p}_{\text{i}\to\text{X}}$ (barn)} & {$\sigma^\text{S1}_{\text{i}\to\text{X}}$ (barn)} & {$\PP{S1}{i}{i}$ (\%)} & {$\frac{\PP{S1}{i}{i}(B)}{\PP{S1}{i}{i}(A)}$ (\%)} \\
        \midrule
         A & \iso{12}{C} & 0.781$\pm$0.021& 0.250$\pm$0.012 \ & 1.056$\pm$0.033& 97.9$\pm$0.1 &  \\
         \midrule[0.06em]
         \multirow{3}{*}{B} & \iso{11}{C} & 0.64$\pm$0.07 & 0.27$\pm$0.06 & 0.94$\pm$0.09 & 97.8$\pm$0.2 & 100.2$\pm$0.2 \\
           & \iso{11}{B} & 0.58$\pm$0.21& 0.61$\pm$0.20  & 1.26$\pm$0.30 & 97.1$\pm$0.7 & 99.5$\pm$0.7 \\
           & \iso{10}{B} & 0.91$\pm$0.15 & 0.52$\pm$0.14 & 1.48$\pm$0.21 & 96.6$\pm$0.5 & 99.0$\pm$0.5 \\
         \bottomrule
    \end{tabular}}
\end{center}
\end{table}

In the case of \iso{11}{C} production, the interaction cross sections of the primary \iso{11}{C} and \iso{12}{C} are required.
Since these nuclei are present in the data, the corresponding in-target mass-changing probabilities are a measured quantity (see \cref{app:A1}).
The computed cross section values and the corresponding survival probabilities of the two nuclei are given in \cref{tab:a51}.
The same technique is applied for the analysis of boron isotopes as well.
The last column of~\cref{tab:a51} gives the ratio of the upstream survival probabilities nucleus $\text{\f}=\{\text{\iso{11}{C}}, \text{\iso{11}{B}}, \text{\iso{10}{B}}\}$ and $\text{\b}=\text{\iso{12}{C}}$, which is very close to 1, hence simplifying the formalism as expressed in \cref{eq:9}

The second term in \cref{eq:9} contains the probability, $\PP{up}{b}{f}$.
It quantifies the fragmentation of the beam nucleus \b into nucleus \f\,upon undergoing an interaction in the upstream region.
Analogous to the explanation with the upstream survival of a nucleus, we associate the upstream production of \f to the interaction b+S1, which can then be calculated by using \cref{eq:A52} for the production cross section of \f.
Alternatively, to determine the ratio, the quantity $\PP{up}{b}{f}$ can be substituted by the total measured production probability of nucleus \f with the PE target, $\PP{PE}{b}{f}$ (calculated and given in \cref{tab:stats}).
Then the ratio term is expressed in terms of known quantities as $\PP{PE}{b}{f}/\PP{S1}{b}{b}$, and is given in \cref{tab:a52}.
The denominator corresponds to upstream survival of \iso{12}{C} and is calculated in \cref{tab:a51}.
\begin{table}[t]
    \caption{The in-target production probability of the \iso{11}{C}, \iso{11}{B}, and \iso{10}{B} nuclei from fragmentation of the \iso{12}{C}.
    \\$^\dag$\,The denominator in this ratio is the same as the $\PP{S1}{i}{i}$ for \iso{12}{C} given in \cref{tab:a51}. }
    \label{tab:a52}
    \begin{center}
    \begin{tabular}{lllr}
    \midrule
        Fragment (f) & {$P^\text{PE}_{^{12}\text{C}\to\text{f}}$} & {$P^\text{S1}_{^{12}\text{C}\to^{12}\text{C}}$ $^\dag$} &  {$\left(\frac{P^\text{S1}_{^{12}\text{C}\to\text{f}}}{P^\text{S1}_{^{12}\text{C}\to^{12}\text{C}}}\right)$} \\ %
         \midrule
         \iso{11}{C} & $0.0060\pm0.0002$ & {\multirow{3}{*}{$0.979\pm0.001$}} & $0.0061\pm0.0002$\\
         \iso{11}{B} & $0.0058\pm0.0006$ &  & $0.0059\pm0.0006$ \\
         \iso{10}{B} & $0.0036\pm0.0002$ &  & $0.0037\pm0.0002$ \\
         \bottomrule
    \end{tabular}
    \end{center}
\end{table}
The ratio of the two quantities is $\mathcal{O}(10^{-3})$.
Moreover, the two target survival probabilities in \cref{eq:9} are calculated using \cref{eq:5}.
Their values are of the same order, that is, $\PP{T}{f}{f} \approx \PP{T}{b}{b}$, for the targets $\text{T}=\{\text{PE}, \text{C}\}$.
Therefore, their difference is ${\sim}\mathcal{O}(10^{-3})$.
The total contribution of the second term in \cref{eq:9} significantly reduced to ${\sim}\mathcal{O}(10^{-5})$.

\section{Calculation of Corrections and Systematic Uncertainties}
\label{app:B}

The cross section values corresponding to mass-changing reactions and specific isotope production are computed as described in \cref{sec:analysis,sec:ccXsec,sec:isoXsec} and are subject to further corrections.
For instance, the feed-down probability due to the mixed composition of the beam affects the true in-target production of our fragment of interest and needs to be corrected.
Such corrections also introduce a systematic uncertainty in our measurement.
The following subsections describe the various corrections applied to our results.
The details including mathematical derivations related to some of the corrections are given in~\cref{appendix}.

\subsection{Feed-down due to Beam Impurities}
\label{sec:imp}

The purity of the beam is defined as the ratio of the number of wanted nuclei, which in this case is \iso{12}{C}, to the total number of particles selected in the upstream cuts.
Hence, other neighboring nuclei like \iso{14}{N}, \iso{15}{N}, \iso{13}{C} etc.\ are the beam impurities that can fragment into the lighter nuclei.
Their feed-down to relevant isotopes of interest like \iso{11}{C} needs to be calculated and corrected in the final computation of the production cross section.
The impact of these beam impurities depends on the number of events in the upstream selection of \iso{12}{C} as described in \cref{app:A4}.
The fractions of these nuclei $f_\text{i}$ relative to \iso{12}{C}, $f_{^{12}\text{C}}$ is given in \cref{tab:A2}.
The true production probability $P^\text{true,T}_{^{12}\text{C}\to\text{f}}$ corrected for the feed-down from beam impurities is given by,
\begin{equation}
P^\text{true,T}_{^{12}\text{C}\to\text{f}} =
  \frac{P^\text{measured,T}_{^{12}\text{C}\to\text{f}} - \sum_\text{i} \Ps{i}{f} \, f_i}
       {f_{^{12}\text{C}}}.
\label{eq:14}
\end{equation}
The expected true probability is expressed in terms of the measured production probability of fragment \f, and subtracting the feed-down from other nuclei, $\Ps{i}{f}$ (see \cref{app:A2} for full derivation).
The feed-down probability is a measured quantity and is calculated by modifying the upstream cuts to select the impurity nuclei as the primary beam particle and measuring the production of the nucleus \f.
The total number corresponding to each impurity nucleus present in the beam composition is given in \cref{app:A1}.
The systematic uncertainty due to this correction is calculated using the statistical uncertainty of the auxiliary feed-down measurements.
The correction and the resulting systematic uncertainty, on each of the calculated production cross sections for the carbon and boron isotopes are given in \cref{tab:corr}.

\subsection{Interaction Inside the Target}
\label{sec:recIntTar}

The target thickness is ${\sim}10\%$ of the nuclear interaction length of the fragment nuclei produced in the beam-target interaction.
Therefore, there is a small yet finite probability, that the nuclear fragments produced inside the target can inelastically interact with the target nucleus.
This interaction results in the underestimation of the number of produced fragment nuclei measured in the MTPC, decreasing its final production cross section value. Therefore, the measured interaction probability corresponds to two consecutive interactions for the reaction sequence, b+T$\to$f \& f+T$\to$X and can be mathematically expressed as a convolution of the two interaction probabilities $\PP{T}{b}{f}$ and $\PP{T}{f}{X}$.

In addition to the measured probability of production of the fragment \f ($\PP{m}{b}{f}$), the two key inputs required for this correction are the measured mass-changing interaction probabilities of the beam \b, $\Ps{b}{b}$ and the fragment nucleus \f, $\Ps{f}{f}$.
These probabilities are measurable quantities and can be computed by altering the upstream selection cuts appropriately.
The final expression for the true production cross section of \f is then written as,
\begin{equation}
\sigma^\text{true,T}_{\text{b}\to\text{f}} =
  \frac{1}
       {n_\text{T}\,d_\text{T}}
  \left(
    \frac{\ln\Ps{f}{f} - \ln\Ps{b}{b}}
         {\Ps{f}{f} - \Ps{b}{b}}
  \right)
  \PP{meas}{b}{f}.
\label{eq:16}
\end{equation}
As in the case with the correction due to feed-down from beam nuclei (\cref{sec:imp}), the systematic uncertainty resulting from this correction is calculated using the statistical uncertainty of the measured mass-changing probabilities of the nuclei \b and \f.
The complete explanation of the derivation of \cref{eq:16} is detailed in \cref{app:A3}.

\subsection{Track Selection in the MTPC}
\label{sec:trackSelMTPC}

The identification of charged fragments in the MTPC is based on the energy deposit \dedx in the detector volume, which is proportional to the squared charge $Z^{2}_\text{MTPC}$. Carbon and boron fragment tracks are selected based on the squared charge cuts as explained in~\cref{sec:downSel}. This results in loss of carbon and boron tracks beyond their respective $Z^{2}_\text{MTPC}$ limits, leading to a selection inefficiency. It can be corrected by performing a binned fit to the 1-dimensional $Z^{2}_{MTPC}$ distribution of the fragments, in the range corresponding to the rising edge of the carbon peak $31.0\leqslant Z^{2}_\text{MTPC} \leqslant 33.0$, for all three target settings, PE, C, and OUT. The fraction of lost tracks is determined from the integral of the function extrapolated beyond the lower limit, that is, in the range $0.0\leqslant Z^{2}_\text{MTPC}\leqslant 31.0$ for the carbon fragments. Therefore, the correction factor $\epsilon_\text{12C}$ is written as:
\begin{equation}
	\epsilon_\mathrm{12C} = \frac{N^\text{m}_\text{12C}}{N^\text{true}_\text{12C}} = \frac{N^\text{m}_\text{12C}}{N^\text{m}_\text{12C} + \Delta{N}_\text{12C}},
	\label{eq:corr1}
\end{equation}
where $N^\text{m}_\text{12C}$ is the measured number of \iso{12}{C} tracks in the range $31.0\leqslant Z^{2}_\text{MTPC} \leqslant 44.0$, while $\Delta{N}_{12C}$ is the extrapolated number determined from the fit. The sum of these two quantities is the total number of \iso{12}{C} in the MTPC denoted as $N^\mathrm{true}_\mathrm{12C}$.

As seen in~\cref{fig:2Dplot}, since \iso{12}{C} and \iso{10}{B} share the same \delx=0.0\,cm based on their deflection in the vertex magnetic field. Therefore, the \iso{12}{C} tracks in the tail of the carbon peak, affect the number of measured boron tracks and alter the final cross section. This is accounted for by estimating the number of \iso{12}{C} in the selection range for boron tracks, $22.0 \leqslant Z^{2}_\text{MTPC} \leqslant 27.5$ from the fit function. The surplus $\Delta{N}_\text{10B}$ must be subtracted from the measured number of boron in the MTPC. Therefore, in this case, the correction factor $\epsilon_\text{10B}$ is written as,
\begin{equation}
	\epsilon_\text{10B} = \frac{N^\text{m}_\text{10B}}{N^\text{true}_\text{10B}} = \frac{N^\text{m}_\text{10B}}{N^\text{m}_\text{10B} - \Delta{N}_\text{10B}},
	\label{eq:corr2}
\end{equation}
The systematic uncertainty due to this correction is calculated from the statistical uncertainty on $\Delta{N}$.

\subsection{Beam Interaction in the Detectors}
\label{sec:reIntDet}

The nuclear fragments produced at the target are subject to spallation at the support structures of the TPCs.
This results in a loss of fragments produced ins the target and introduces a systematic uncertainty.
The computed production cross section needs to be corrected for this loss, nevertheless, it is implicit in the formalism adopted in this work (detailed in \cref{sec:isoXsec}), and needs no additional compensation.
The product of the probabilities in the second term of \cref{eq:4}, $\PP{up}{b}{b}\,\PP{T}{b}{f}\,\PP{dn}{f}{f}$, ensures that fragment tracks measured in the MTPC-L are produced by beam particle spallation at the target and survive any inelastic interaction in the detectors and its support structures downstream of the target.

For instance, in the case of \iso{11}{C} production, the auxiliary measurements made with \iso{11}{C} as the primary beam particle are used to calculate the total target out mass-changing probability, which is further input to \cref{eq:8} to calculate the final production probability of the isotope from the beam-target interaction.

\begin{table}
\caption{A summary of the calculated corrections ($\Delta\sigma$) applied to the measured isotope production cross sections in \iso{12}{C}+p interactions, and their corresponding systematic uncertainty ($\updelta\sigma$).}
\label{tab:corr}
\begin{center}
\begin{tabular}{l c c c c c c}
\toprule
\multirow{2}{*}{Correction} & \multicolumn{2}{c}{\iso{12}{C}+p$\to$\iso{11}{C}} & \multicolumn{2}{c}{\iso{12}{C}+p$\to$\iso{11}{B}} & \multicolumn{2}{c}{\iso{12}{C}+p$\to$\iso{10}{B}}\\
\cline{2-7}
& $\Delta\sigma$ (mb) & $\updelta\sigma$ (mb) & $\Delta\sigma$ (mb) & $\updelta\sigma$ (mb) & $\Delta\sigma$ (mb) & $\updelta\sigma$ (mb) \\
\midrule
Beam impurities & $+0.09$ & $\pm0.04$ & $+0.02$ & $\pm0.02$ & $+0.01$ & $\pm0.02$ \\
In-target re-interaction & $+1.98$ & $\pm0.14$ & $+3.26$ & $\pm0.37$ & $+1.42$ & $\pm0.12$\\
Track Selection & - & - & - & - & $+0.01$ & $\pm0.04$ \\
Target density & - & $\pm0.05$ & - & $\pm0.06$ & - & $\pm0.03$ \\
\midrule
Total & $+2.08$ & $\pm0.23$ & $+3.28$ & $\pm0.45$ & $+1.44$ & $\pm0.21$ \\
\bottomrule
\end{tabular}
\end{center}
\end{table}

\subsection{Target Density Uncertainty}
\label{sec:tdu}

Another source of systematic uncertainty is the uncertainty in the measured density of both the targets, PE and C.
The graphite of the C target is the same as that of the T2K target with a known mass density of $\rho_\text{C} = 1.84\pm0.01$\,g/cm$^3$.
Similarly, for the polyethylene target (PE), with density $\rho_\text{PE} = 0.93$\,g/cm$^3$, the estimated uncertainty in the density from dimensional measurements is also ${\sim}1\%$ of the specified mass density of the material.
The systematic uncertainty on the measured cross sections is then estimated by propagating these uncertainties in the calculation.
Its contribution is approximately 15\% of the total calculated systematic uncertainty.

\end{document}